\documentclass[aps,pre,twocolumn,showpacs,final,floatfix]{revtex4-2}
\usepackage{amsmath}
\usepackage{amsfonts}
\usepackage{amssymb}
\usepackage{physics}
\usepackage{color}
\usepackage{hyperref}
\usepackage{inputenc}
\usepackage{enumerate}
\usepackage{graphicx}
\usepackage{float}
\usepackage{amsmath}
\usepackage{amsfonts}
\usepackage{lipsum}
\usepackage{dcolumn}
\usepackage{hyperref}
\usepackage{subfigure}
\usepackage{epsfig}
\usepackage{epsf} 
\usepackage{epstopdf}

\usepackage{color}
\usepackage{enumitem}
\usepackage{xr}
\makeatletter
\DeclareMathOperator{\sign}{sign}

\newcommand*{\addFileDependency}[1]{% argument=file name and extension
\typeout{(#1)}% latexmk will find this if $recorder=0
\@addtofilelist{#1}
\IfFileExists{#1}{}{\typeout{No file #1.}}
}\makeatother

\usepackage{hyperref}
\hypersetup{colorlinks,
    citecolor=blue,
    filecolor=cyan,
    linkcolor=blue,
   urlcolor=magenta
 }

\usepackage{cleveref}

\begin{document}

\title{Yang-Lee Zeros of Certain Antiferromagnetic Models}

% Please use the appropriate macro for each type of information

\author{Muhammad Sedik$^1$}
\author{Junaid Majeed Bhat$^1$}

%\thanks{}
\author{Abhishek Dhar$^1$}
\author{ B Sriram Shastry$^1$}
\affiliation{$^1$University of California, Santa Cruz, CA, 95064, \\ 
$^2$International Centre for Theoretical Sciences,  Tata Institute of Fundamental Research, Bengaluru 560 089, India.}

%Tata Institute of Fundamental Research,
%Collaboration name if desired We present a stochastic formulation of the Kadanoff-Baym or Keldysh theory to calculate the conductance


\date{\today}

\begin{abstract}
 We revisit the somewhat less studied problem of  Yang-Lee zeros of the Ising antiferromagnet. For this purpose, we study two models, the nearest-neighbor model on a square lattice, and the more tractable mean-field model corresponding to infinite-ranged coupling between all sites. In the high-temperature limit, we show that the logarithm of the Yang-Lee zeros can be written as a series in half odd integer powers of the inverse temperature, $k$,  with the leading term  $\sim k^{1/2}$. This result is true in any dimension and for 
 arbitrary lattices. We also show that the coefficients of the expansion satisfy simple identities (akin to sum rules) for the nearest-neighbor case. These new identities are verified numerically by computing the exact partition function for a 2D square lattice of size $16\times16$. For the mean-field model, we write down the partition function (termed the mean-field polynomials) for the ferromagnetic (FM) and antiferromagnetic (AFM) cases and derive from them the mean-field equations. We analytically show that at high temperatures the zeros of the AFM mean-field polynomial scale as $\sim k^{1/2}$ as well. Using a simple numerical method, we find the roots lie on certain curves (the root curves), in the thermodynamic limit for the mean-field polynomials for the AFM case as well as for the FM one.  Our results show a new root curve, that was not found earlier. Our results also clearly illustrate the phase transition expected for the FM and AFM cases, in the language of Yang-Lee zeros. Moreover, for the AFM case,  we observe that the root curves separate two distinct phases of zero and non-zero complex staggered magnetization, and thus depict a complex phase boundary.
  
\end{abstract}

% insert suggested PACS numbers in braces on the next line
%\keywords{}
\maketitle

%\tableofcontents
\section{Introduction}

In the two seminal papers in 1952~\cite{yang1952statistical,lee1952statistical}, Yang and Lee introduced a new method to study the phase transition of models in statistical physics by studying the distribution of the zeros of the partition function in the complex fugacity plane. In particular, the behavior of these Yang-Lee zeros near the positive real axis describes the system properties near phase transitions. They also proved the famous Yang-Lee theorem for the ferromagnetic case, which states that the partition function of the ferromagnetic (FM) Ising model in an external magnetic field $h$ has zeros only on the unit circle in the complex $e^{-2\beta h}$-plane. Yang-Lee theorem was also extended to various other ferromagnetic type models~\cite{ruelle1971extension,griffiths1973phi,newman1974zeros,biskup2000general,lieb1981general}, and to study non-equilibrium phase transition~\cite{arndt2000yang}. The Yang-Lee zeros of the FM Ising model were also studied analytically and numerically on different lattices~\cite{kim2023exact,kim2020yang,deger2020lee,kim2018partition,ghulghazaryan2002yang}. The distribution of the Yang-Lee zeros on the unit circle is also of interest since it can be probed experimentally. The earliest attempt used experimental data for magnetization of a two-dimensional Ising ferromagnet to determine this  distribution~\cite{binek1998density}. Other experimental realizations of the Yang-Lee zeros are presented in~\cite{wei2012lee,peng2015experimental,brandner2017experimental}. A full review of Yang-Lee formalism and its applications can be found in~\cite{bena2005statistical}. 

Turning to the antiferromagnetic (AFM) case, the behavior of the zeros for  AFM interaction on two and higher-dimensional lattices is much less understood compared to the FM case. 
The main difficulty in applying popular numerical techniques, such as the Monte-Carlo method to this problem, is the extreme sensitivity of the roots to the numerical precision of the computation. For systems as big as the ones studied here, one needs the partition function with essentially exact (or infinite precision) arithmetic, making the problem quite hard.   One motivation for studying the antiferromagnetic Ising model is that it may be viewed as an example of a  classical lattice gas with repulsive interactions. A clear understanding of the location of partition function zeros in this system is a prerequisite for studying topical and important quantum lattice gas models, such as the repulsive Hubbard model. 

The 1-d model was solved already by Lee-Yang~\cite{lee1952statistical} for either sign of the exchange, but since the model does not have a finite temperature phase transition, the case of higher dimensions remained unclear and interesting.
The AFM Ising on a two-dimensional lattice was studied for small system sizes in~\cite{suzuki1970statistical,katsura1971distribution}, and the zeros were found to lie in the negative half of the complex $e^{-2\beta h}$-plane. However, it is known that this model undergoes a disorder phase transition at critical magnetic field $h_c$ for low enough temperatures~\cite{dobrushin1985phase,dasgupta1982phase}. Therefore, the zeros must jump to the positive half-plane and touch the real axis in the thermodynamic limit so that this phase transition is realized.  These features were observed by Kim by evaluating the Yang-Lee zeros of the Ising model on a square lattice numerically for sizes up to $14\times14$~\cite{kim2004yang}. Although Yang-Lee zeros of this somewhat small system are roughly consistent with the expectations of the phase diagram, the sparse zeros don't give enough information about a root curve or in general the locus of zeros in the thermodynamic limit. A numerical study of the Yang-Lee zeros of the AFM Ising model on the triangular lattice can be found in~\cite{hwang2010yang}. Some analytical work was done on other AFM models like the anisotropic Heisenberg chain~\cite{katsura1962statistical} and other models~\cite{lebowitz2012location,heilmann1971location}. However, the analytical work on AFM nearest-neighbor Ising on a two-dimensional lattice is limited. Some examples of such works are Lieb and Ruelle showing that there are regions free of zeros for temperatures above the critical temperature~\cite{lieb1972property}, and Heilmann and Lieb proving that all the zeros lie on the negative real axis for high enough temperatures~\cite{heilmann1970monomers}.

In this work, we attempt to arrive at a better understanding of the Yang-Lee zeros for the antiferromagnetic Ising model. We first consider the nearest neighbor Ising model and obtain an expansion for the location of the (logarithm of) zeros in terms of the inverse temperature, $k$ in the high-temperature limit. We show that irrespective of the dimensions and the geometry of the lattice, the leading term in the expansion goes as $\sim \sqrt{k}$. We also show that the coefficients of the expansion follow simple identities (i.e. sum-rules) which we verify by numerically evaluating the partition function and thereby the zeros of the Ising model on a square lattice of size $16\times16$. 

Given the difficulty of finding the locus of Yang-Lee zeros of the AFM Ising model in the thermodynamic limit, we constructed \textit{mean-field (MF) polynomials}, $\mathcal{Z}_{\text{FM}}$ and $\mathcal{Z}_{\text{AFM}}$, whose zeros behave similarly to the zeros of the partition function of the FM and AFM Ising model on a square lattice respectively. We realized after completion of this work that similar polynomials, which describe the partition function of Bragg-Williams approximation of the lattice gas, were introduced and studied in~\cite{katsura1954phase,ohminami1972distribution}. It can be shown that the FM polynomial has zeros on the unit circle in agreement with the Yang-Lee theorem. One can also provide an analytical derivation
of the density of the zeros for temperatures below the critical temperature. For the AFM case, Ohminami et al developed a criterion for where the zeros can occur in the thermodynamic limit~\cite{ohminami1972distribution}. They showed that the zeros lie on the boundary that separates two regions in the z-plane such that one region is a complex paramagnetic phase, whereas the other is a complex antiferromagnetic phase. They showed the root curves in the thermodynamic limit by a numerical search for points where the criterion is satisfied. 
 
 The present work extends these studies in several directions. We prove that $\mathcal{Z}_{\text{AFM}}$ at high temperatures is a linear combination of Hermite polynomials. The roots of this linear combination scale as a power of $\sqrt{k}$. We then present a technique that uses the free energy to compute the density of Yang-Lee zeros in the thermodynamic limit and show its numerical results on the MF polynomials. This technique provides a more extensive search for the roots. Such extensive search leads to finding new root curves that were not shown before in literature. Finally, we show that those new root curves satisfy the criteria developed by Ohminami et al in~\cite{ohminami1972distribution}.

This paper is structured as follows: in Sec.~\ref{sec:Ising Model}, we study the Ising model on arbitrary lattice at high temperatures and show that the logarithm of Yang-Lee zeros is a power series in half odd integer powers of the inverse temperature $k$, with leading term $\sqrt{k}$. We also derive two new identities (i.e. sum rules) that the power series coefficients satisfy. We conclude the section by numerically verifying the sum rules for the nearest-neighbor Ising model on a square lattice of size $(16\times 16)$. In Sec.~\ref{sec:MFPoly}, we introduce the mean-field model and write down the partition functions for FM and AFM interaction. At high temperatures, we show that the logarithm of Yang-Lee zeros scale as $~k^{1/2}$ using two different approaches and that the AFM mean-field polynomial is a linear combination of Hermite polynomial. We numerically find the roots in the thermodynamic limit using a simple technique involving the free energy per site. The results of this technique show root curves that were not presented before in literature, to our best knowledge. We finally discuss our results of the roots in the thermodynamic limit and interpret them as a phase boundary in the complex $z$-plane.

\section{Ising Model}
\label{sec:Ising Model}

Consider the Ising Hamiltonian
\begin{equation}
     \mathcal{H}=J\sum_{\langle ij \rangle}(1-\sigma_i\sigma_j)-h\sum_{i}(1+\sigma_{i}),
    \label{H1}
\end{equation}
where $\sigma_i=\pm 1$ is the $i^{\text{th}}$ spin on an arbitrary regular lattice with number of sites $N_s$ and number of bonds $N_b$, which depends on the chosen boundary conditions. $J$ is the coupling constant between spins, and $h$ is an applied external magnetic field. $\langle ij \rangle$ indicates sum over nearest neighbors. The partition function for this model is given by,
\begin{equation}
        \mathcal{Z}=\sum_{\{\boldsymbol{\sigma}\}}e^{-\beta \mathcal{H}}=\sum_{n_b=0}^{N_b}\sum_{n_s=0}^{N_s}\Omega(n_b,n_s)u^{n_b}z^{n_s},
        \label{Z(omega)}
\end{equation}
where  $u=e^{-2\beta J}$, $z=e^{2\beta h}$ and  $\beta= \frac{1}{k_B T}$. $\Omega(n_b,n_s)$ is the number of states with interaction energy $E$ and magnetization $M$, which are given by
\begin{equation}
        E=\frac{J}{2}\sum_{\langle ij \rangle}(1-\sigma_i\sigma_j)=n_b,  ~ M=\frac{1}{2}\sum_{i}(1+\sigma_i)=n_s.
    \label{IntEnergy/Mag}
\end{equation}
At a fixed $u$, the partition function $\mathcal{Z}$ is a polynomial of degree $N_s$ in the variable $z$. The roots of this polynomial are the so-called Yang-Lee zeros. The Yang-Lee zeros completely specify the thermodynamic state of the system. Consequently, their behavior around the real axis, in the complex $z$-plane, describes the phase transitions, if any, of the system. The phase transition is characterized by the Yang-Lee zeros touching the real axis in the thermodynamic limit, which results in non-analytic free energy, $\beta f=\lim_{N_s\rightarrow\infty}(1/N_s)\ln(\mathcal{Z})$.

Since the Ising model admits $h\rightarrow -h$ symmetry, which is reflected in $z \rightarrow 1/z$, it suffices to consider only the roots within the unit disk $|z|\leq 1$. We focus on the AFM interaction case $(J<0)$ because the distribution of the zeros at any arbitrary temperature is not known. However,  it is straightforward to see, from Eq.~(\ref{Z(omega)}), the behavior of the roots, for the two extreme limits of the temperature. At zero temperature $(u = 0)$, the roots in the interior of the unit circle lie at $z=0$. At infinite temperature $(u=1)$, there is a single root at $z=-1$ with multiplicity $N_s$. The behavior of the roots at such high temperatures could be understood through high-temperature expansion, which we consider in detail in the next section. 

\subsection{Yang-Lee Zeros At High Temperature}
\label{subsec:HighTemp}

Let the roots of the partition function be ${z_j}$, where $j=\{1,2,\dots,N_s\}$ at $k=\beta J$. We define the variable 
\begin{equation}
    \xi=-\frac{1}{2}\ln{(-z)}.
    \label{DefineXi}
\end{equation}
The set $\{z_j\}$ is transformed uniquely to $\{\xi_j\}$ if we restrict $-\pi < \arg(z) \leq \pi$. %The symmetry in the new variable is then $\xi_j \rightarrow -\xi_j$. 
For the ferromagnetic and antiferromagnetic interaction,  $z_j \rightarrow -1 \implies \xi_j \rightarrow 0$ for all $j$ in the infinite temperature limit. In terms of the zeros the partition function is given by $\mathcal{Z}=\prod_{j=1}^{N_s}(z-z_j)$, and therefore, expanding at high temperatures, we find
\begin{align}
    \frac{1}{N_s}\ln{\mathcal{Z}}&=\frac{1}{N_s}\sum_{j=1}^{N_s}\ln{(z+e^{-2\xi_j})}
    \notag\\
    &=\ln{(1+z)}+\frac{2z}{(1+z)^2} \langle \xi_j^2 \rangle_j\notag\\
    &~+\frac{2(z-4z^2+z^3)}{3(1+z)^4}\langle \xi_j^4 \rangle_j+\mathcal{O}\left(\langle \xi_j^6 \rangle_j\right), 
    \label{PolyExpand}
\end{align}
%+\frac{(z-4z^2+z^3)}{24(1+z)^4}\langle \xi_j^4 \rangle_j
where $\langle f_j \rangle_j = \frac{1}{N_s}\sum_{j=1}^{N_s}f_j$. The function $\ln{(z+e^{-2\xi_j})}$ was expanded as a Taylor series around $\xi_j=0$ and the odd order terms are canceled after carrying out the summation because $\pm \xi_j$ are both in the set $\{\xi_j\}$. The expansion in the above equation should be the same as the direct high-temperature expansion from Eq.~(\ref{Z(omega)})~\cite{oitmaa2006series} which depends on the lattice and the boundary conditions.  Let us consider the simplest case for periodic boundary conditions on a regular lattice of coordination number $\gamma$. For this case, the direct high-temperature expansion is given by
\begin{align}
\frac{1}{N_s}\ln{\mathcal{Z}}&=\ln{(1+z)}-\frac{N_b}{N_s}\frac{4z}{(1+z)^2} k \notag\\
&~+ 4\frac{N_b}{N_s}
     \frac{z(2z+\gamma(z-1)^2)}{(1+z)^4}k^2+\mathcal{O}(k^3).\label{HighTExpand}
\end{align} 
It is possible to deduce information about the zeros by comparing Eq.~(\ref{PolyExpand}) and Eq.~(\ref{HighTExpand}). Note that the former series is in terms of  $\xi_j$ while the latter is a power series in $k$. Therefore, to compare the two we assume an expansion of $\xi_j$ around $k=0$  as follows,
\begin{equation}
    \xi_j=c_{j0} k^{\alpha_0}+c_{j1} k^{\alpha_1}+c_{j2} k^{\alpha_2}+\dots,
    \label{Xiofk}
\end{equation}
where $\alpha_0<\alpha_1<\alpha_2<\dots$  and $c_{j\nu}$ are arbitrary coefficients, with $\nu=0,1,2,\dots$. For the AFM case ($k<0$), we restrict $c_{j\nu}$ to be real numbers and the power series in $|k|$ as the AFM Yang-Lee zeros at high temperature are all real and negative ~\cite{heilmann1970monomers}.  The lowest order of $|k|$ in Eq.~(\ref{HighTExpand}) is the first order, and the lowest order after plugging Eq.~(\ref{Xiofk}) into Eq.~(\ref{PolyExpand}) is $|k|^{2\alpha_0}$, therefore $\alpha_0=\frac{1}{2}$. All the powers of $|k|$  in Eq.~(\ref{HighTExpand}) are integers. However, due to the $\xi_j^2$ in Eq.~(\ref{PolyExpand}), powers of the form $|k|^{\frac{1}{2}+\alpha_{\nu}}$ appear. Therefore, all $\alpha_{\nu}$ are half odd integers, which gives the behaviour of the logarithm of AFM Yang-Lee zeros at high temperature as 
\begin{equation}
    \xi_j = \sqrt{|k|}\left(\sum_{\nu=0}^{\infty}c_{j\nu}|k|^{\nu}\right).
    \label{Xibehave}
\end{equation}
The above result is due to the presence of the linear term and the integer powers of $|k|$ in Eq.~(\ref{HighTExpand}). These two do not require the lattice to be periodic, and therefore, this result holds for any boundary condition. Moreover, the coefficients $c_{j\nu}$ are not independent of one another. The expansions in Eq.~(\ref{PolyExpand}) and Eq.~(\ref{HighTExpand}) should match for all $z$, and comparing them gives a set of identities satisfied by $c_{j\nu}$. By substituting Eq.~(\ref{Xibehave}) into Eq.~(\ref{PolyExpand}), we get the high temperature expansion in terms of $|k|$ as 
\begin{align}
    \frac{1}{N_s}&\ln{\mathcal{Z}}=\ln{(1+z)}+\frac{2z}{(1+z)^2} \mathcal{A} |k|\notag \\
    &+\left( \frac{4z}{(1+z)^2} \mathcal{B}+\frac{2(z-4z^2+z^3)}{3(1+z)^4}\mathcal{C}\right) |k|^2+\mathcal{O}\left(|k|^3\right), 
    \label{PolyExpandk}
\end{align}
where $\mathcal{A}=\frac{1}{N_s}\sum_{j=1}^{N_s}c_{j0}^2,~ \mathcal{B}=\frac{1}{N_s}\sum_{j=1}^{N_s}c_{j0}c_{j1},$ and $\mathcal{C}=\frac{1}{N_s}\sum_{j=1}^{N_s}c_{j0}^4$. 
Comparing the $k$-term in the above equation with Eq.~(\ref{HighTExpand}) gives the sum-rule
\begin{equation}
    \mathcal{A}=\frac{1}{N_s}\sum_{j=1}^{N_s}c_{j0}^2=2\frac{N_b}{N_s},
    \label{Sum-Rule1}
\end{equation}
while the $k^2$-term gives \begin{equation}
    \frac{4z}{(1+z)^2}\mathcal{B}+\frac{2(z-4z^2+z^3)}{3(1+z)^4}\mathcal{C}=4\frac{N_b}{N_s}
     \frac{z(2z+\gamma(z-1)^2)}{(1+z)^4}.
    \label{k2Compare}
\end{equation}
After simplifying the above equation, and matching the coefficients of $z,z^2$ and $z^3$ on both sides of the equation, we get two independent linear equations for $\mathcal{B}$ and $\mathcal{C}$, which have the solution 
\begin{align}
        &\mathcal{B}=\frac{1}{N_s}\sum_{j=1}^{N_s}c_{j0}c_{j1}=\frac{1}{3N_s}(\gamma+1)N_b,\notag\\ &\mathcal{C}=\frac{1}{N_s}\sum_{j=1}^{N_s}c_{j0}^4=\frac{2}{N_s}(2\gamma-1)N_b.
    \label{Sum-Rule2}
\end{align}
With similar algebra, similar sum-rules can be derived for any other boundary condition chosen for the lattice. For example, the sum-rules for different boundary conditions on a square lattice are summarized in Table \ref{tab:k2sum-rules}.
\begin{table}[t!]
    \centering
    \begin{tabular}{|c||c|c|}
    \hline
        Boundary Conditions &  $\mathcal{B}$& $\mathcal{C}$\\
        \hline
        \hline
         Periodic & $\frac{1}{3N_s} (5N_b)$&$\frac{2}{N_s} (7N_b)$\\ 
         Cylindrical & $\frac{1}{3N_s} (5N_b-24L)$&$\frac{2}{N_s} (7N_b-48L)$\\ 
         Open & $\frac{1}{3N_s} (5N_b-6L+4)$&$\frac{2}{N_s} (7N_b-12L+8)$\\ 
    \hline
    \end{tabular}
    \caption{$\mathcal{B}$ and $\mathcal{C}$ sum-rules for the $k^2$ term for different boundary conditions on an $L\times L$ square lattice where $\gamma=4$. }
    \label{tab:k2sum-rules}
\end{table}

The dependence of the sum rules on the boundary conditions and coordination number can be avoided by constructing a suitable linear combination of $\mathcal{A}$, $\mathcal{B}$ and $\mathcal{C}$.
Such linear combinations are generated by comparing  Eq.~(\ref{PolyExpand}) and Eq.~(\ref{HighTExpand})  at $z=1$ because at this value of $z$, the direct high-temperature expansion for different boundary conditions becomes of the same form as in Eq.~(\ref{HighTExpand}) up to second order. In addition, up to the second order, at $z=1$ the factors of $\gamma$ also vanish. At $z=1$, Eq.~(\ref{HighTExpand}) gives
\begin{align}
    \frac{1}{N_s}\ln\mathcal{Z}=\ln{(2)}+\frac{N_b}{N_s}|k|+\frac{N_b}{2N_s}|k|^2+\mathcal{O}(|k|^3),
    \label{HighTExpandz1}
\end{align} 
whereas evaluating Eq.~(\ref{PolyExpandk}) at $z=1$ gives the power series
\begin{equation}
    \frac{1}{N_s}\ln\mathcal{Z}=\ln(2)+\mathcal{J}_1 |k|+ \mathcal{J}_2 |k|^2+\dots.
    \label{SumruleExpand}
\end{equation}
 where $\mathcal{J}_\mu$ are coefficients that depend on $c_{j\nu}$. The coefficients of the different powers of $|k|$ should match between Eq.~(\ref{HighTExpandz1}) and Eq.~(\ref{SumruleExpand}). The first two coefficients give the following two sum-rules 
\begin{equation}
    \mathcal{J}_1=\frac{1}{2}\mathcal{A}=\frac{N_b}{N_s},~\text{and}~\mathcal{J}_2=\mathcal{B}-\frac{1}{12}\mathcal{C}=\frac{N_b}{2N_s}.
    \label{SumRules}
\end{equation} 
These sum-rules are independent of the geometry and dimension of the lattice. 

To verify the sum-rules in Eq.~(\ref{SumRules}), we first consider the simple case of the 1-d nearest-neighbor Ising model with periodic boundary conditions. The Yang-Lee zeros of this model are found exactly through the eigenvalues of the transfer matrix and given by 
\begin{align}
     &\xi_j=-\frac{1}{2}\ln\left(\phi_j-s_j\sqrt{\phi_j^2-1}\right)\notag\\
     &\quad\sim c_{j0} \sqrt{|k|}+c_{j1}|k|^{3/2}+\mathcal{O}(|k|^{5/2})
    \label{1DRoots}
\end{align}
where $\phi_j=(e^{-4k}-1)\cos{q_j}+e^{-4k}$, $q_j=\frac{(2j+1)\pi}{N_s}$, $s_j=\sign(2j-N_s+1)$, $ c_{j0}= \sqrt{2}s_j\sqrt{1+\cos{q_j}}$, $c_{j1}=-\frac{2}{3}s_j\left|\cos{\frac{q_j}{2}}\right|\left(\cos{q_j}-2\right)$, and $k<0$ for AFM interaction. By plugging these expressions for $c_{j0}$ and $c_{j1}$ into Eqs.~(\ref{SumRules}), it can be verified that the sum-rules are satisfied with $N_s=N_b$ due to the periodic boundary conditions.

In the next subsection, we verify the sum-rules in Eq.~(\ref{SumRules}) for the 2-d nearest-neighbor Ising model on a square lattice by computing the exact partition function of the $16\times 16$ lattice and finding its zeros exactly. We also describe the behavior of Yang-Lee zeros at different temperatures.

\subsection{Numerical Results On Square Lattice}
\label{Subsec:16x16results}

The sum rules can be verified if the roots are known at different temperatures. We find the roots by computing the partition function polynomial directly. A memory-efficient algorithm to compute the zero-field partition function of some discrete systems was given by Bhanot~\cite{bhanot1990numerical}. We extend this algorithm to finite-field partition functions and compute the 2-d nearest-neighbor Ising model exact partition function for sizes $15\times 15$ and $16\times16$ by calculating the number of states $\Omega(n_b,n_s)$ for open and cylindrical (i.e. periodic in one direction) boundary conditions of the square lattice. The exact algorithm and its extension are explained in appendix \ref{Appx:Bhanot}.

\begin{figure}[b!]
    \centering
    \subfigure[$~T=8 Tc$]{
    \includegraphics[width=0.23\textwidth]{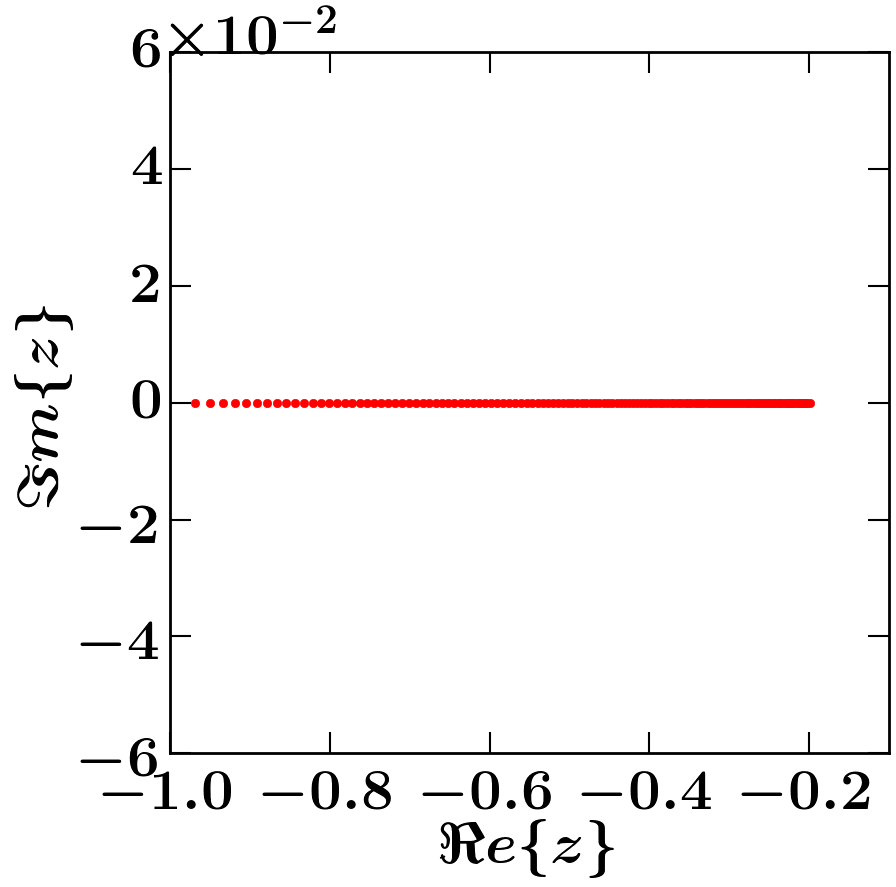}}
    \subfigure[$~T= 2 Tc$]{
    \includegraphics[width=0.23\textwidth]{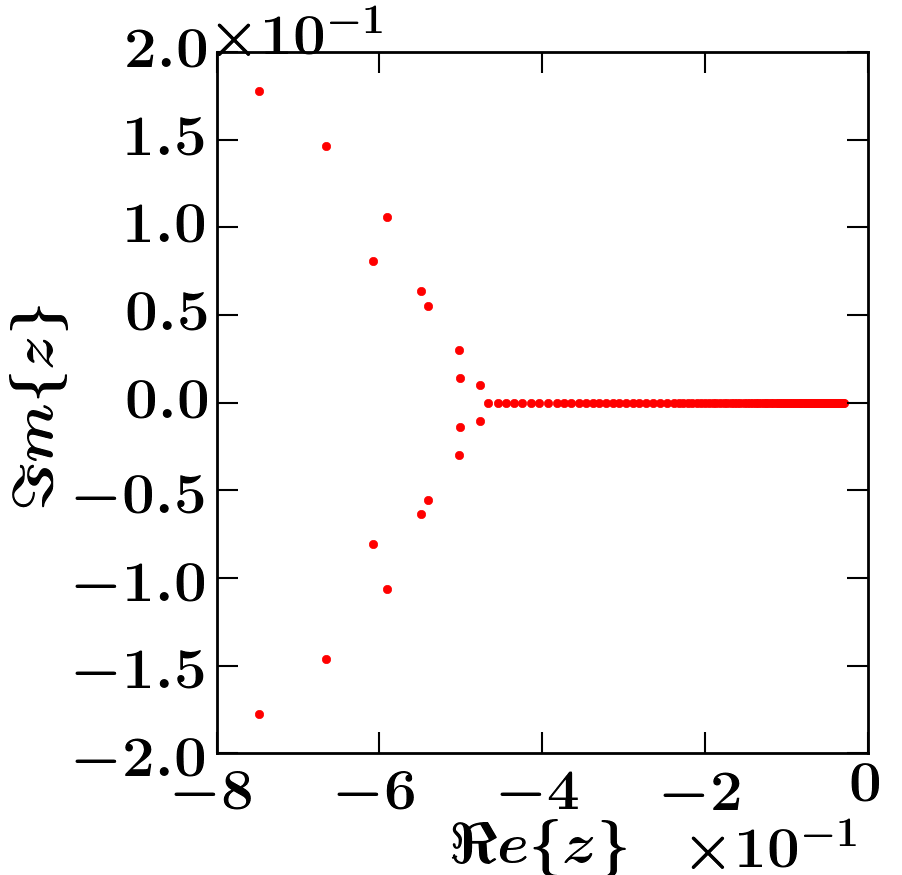}}
    \subfigure[$~T=Tc$]{
    \includegraphics[width=0.23\textwidth]{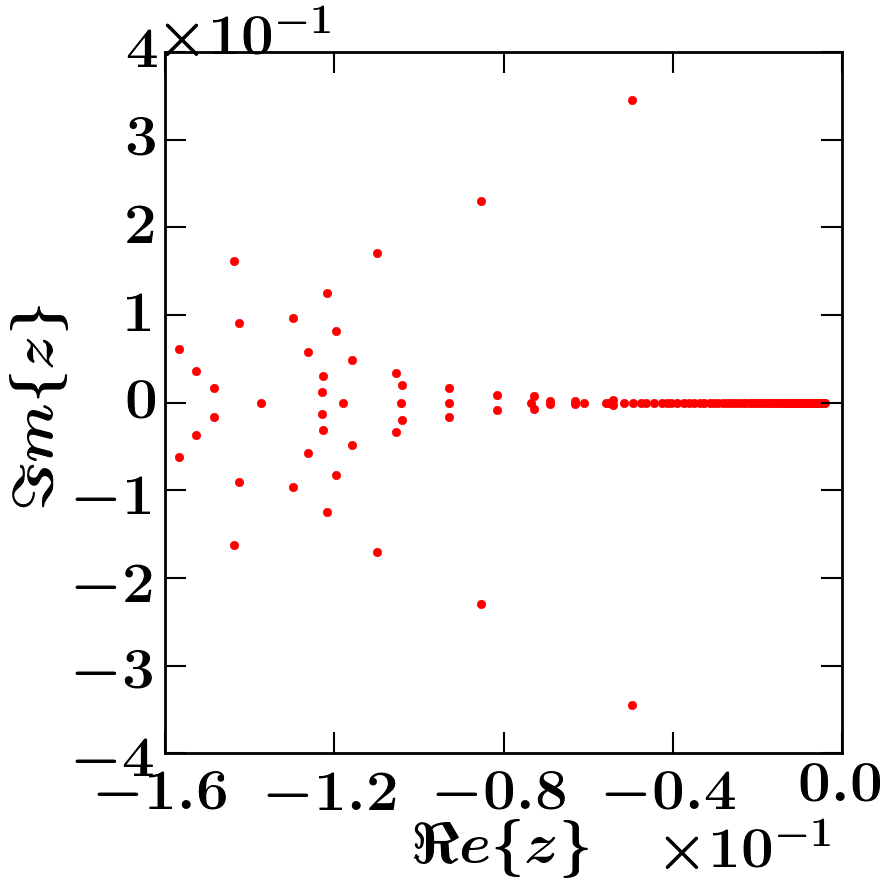}}
     \subfigure[$~T=0.25Tc$]{
    \includegraphics[width=0.23\textwidth]{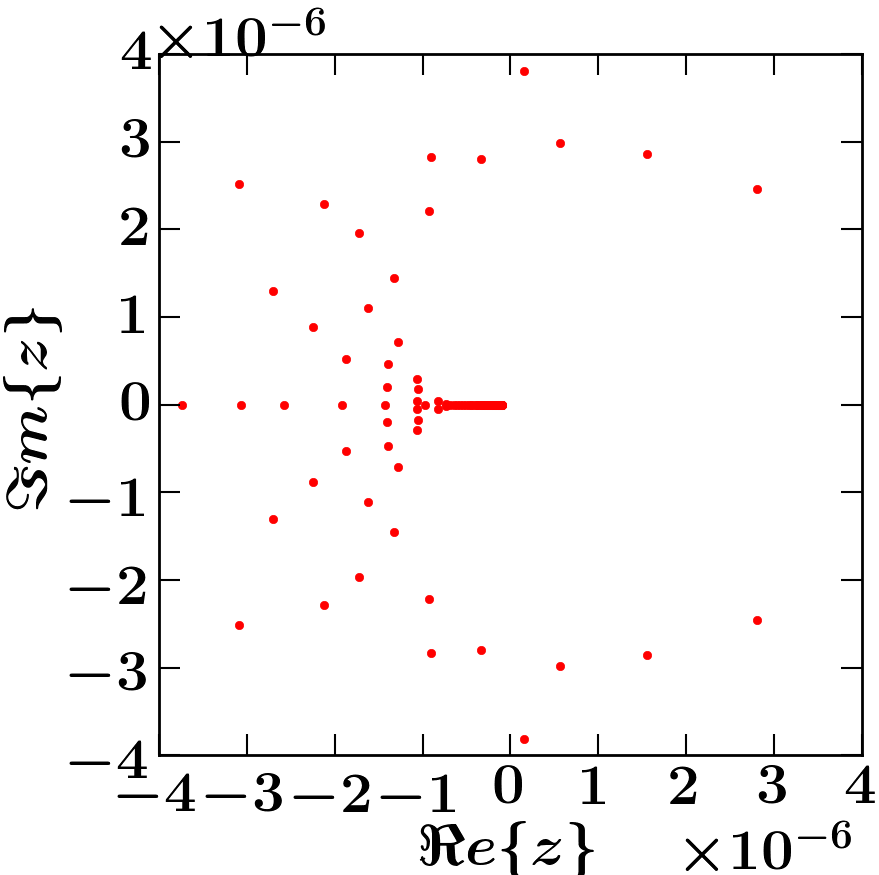}}
    \caption{Yang-Lee zeros in the complex $z$-plane of $16\times 16$ lattice with open boundary conditions. (a)-(d) show different temperatures. As the temperature decreases, roots fly off the negative real axis and enter the positive real half-plane. Furthermore, the roots cluster around $z=0$ at very low temperatures. $T_c=\frac{2}{\ln{(1+\sqrt{2})}}\frac{J}{k_B}$, where $k_B$ is Boltzmann's constant.}
    \label{fig:IsingRoots}
\end{figure}

The Yang-Lee roots of $16\times 16$ lattice with open boundary conditions in the complex $z$-plane inside the unit disk $|z|\leq 1$  for different temperatures are shown in Fig.~\ref{fig:IsingRoots}. At very high temperature $T\gg T_c=\frac{2}{\ln{(1+\sqrt{2})}}\frac{J}{k_B}$, we see all the roots lie on the negative real axis. As the temperature is lowered towards the critical temperature $T_c$, complex roots start to appear. Further decreasing the temperature below $T_c$, some roots jump to the positive real half-plane. The imaginary parts of these roots are expected to decrease with increasing the system size and touch the real axis in the thermodynamic limit. At very small temperatures, the roots cluster around $z=0$ as expected.

%\subsection{Precision Problem}

We now verify the sum-rules in Eq.~(\ref{SumRules}) by computing the left-hand side of 
\begin{equation}
    \frac{1}{2}\langle\xi_j^2\rangle-\frac{1}{12}\langle\xi_j^4\rangle=\mathcal{J}_1 k + \mathcal{J}_2 k^2+\dots
\end{equation}
at different values of $k$, and terminating the series on the right-hand side up to order $n$. We construct an $n\times n$ linear system to numerically find $\{\mathcal{J}_1, \mathcal{J}_2,\dots ,\mathcal{J}_n\}$. We can then confirm the sum-rules for $\mathcal{J}_1$ and $\mathcal{J}_2$ for the 2-d Ising model on $16 \times 16$ square lattice with open boundary conditions. We show this comparison in Table \ref{tab:2dSumrule}, in which $N_s=16^2$ and $N_b=2(16)(16-1)$.

\begin{table}[t!]
    \centering
$$\begin{array}{|c|c|c|c|c|}
\hline
 n & \mathcal{J}_1 & \text{$\%$ error in }\mathcal{J}_1 & \mathcal{J}_2 & \text{$\%$
   error in }\mathcal{J}_2 \\
   \hline
   \hline
 2 & 1.88 & 1\times 10^{-3} & 0.918 & 2 \\
 10 & 1.88 & 2\times 10^{-23} & 0.938 & 1\times 10^{-19}
   \\
 20& 1.88 & 1\times 10^{-43} & 0.938 & 1\times 10^{-39}
   \\
 30 & 1.88 & 2\times 10^{-59} & 0.938 & 2\times 10^{-55}
   \\
 40 & 1.88 & 4\times 10^{-74} & 0.938 & 5\times 10^{-70}
   \\
 50 & 1.88 & 8\times 10^{-89} & 0.938 & 1\times 10^{-84}
   \\
   \hline
\end{array}$$
    \caption{\% error in both sum-rules in Eq.~(\ref{SumRules}) for $\mathcal{J}_1$ and $\mathcal{J}_2$ on a $16 \times 16$ lattice with open boundary conditions. The \% error is computed relative to the high-temperature expansion coefficient in Eq.~(\ref{HighTExpand}) as $\frac{|\mathcal{J}_1-N_b/N_2|}{N_b/N_s}$ and $\frac{|\mathcal{J}_2-N_b/(2N_2)|}{N_b/(2N_s)}$. The values of $k$ used to evaluate this are $-0.02\leq k \leq -0.002$.} 
    \label{tab:2dSumrule}
\end{table}

\begin{figure}[t!]
    \centering
    \subfigure[$~T=2 Tc$]{
    \includegraphics[width=0.23\textwidth]{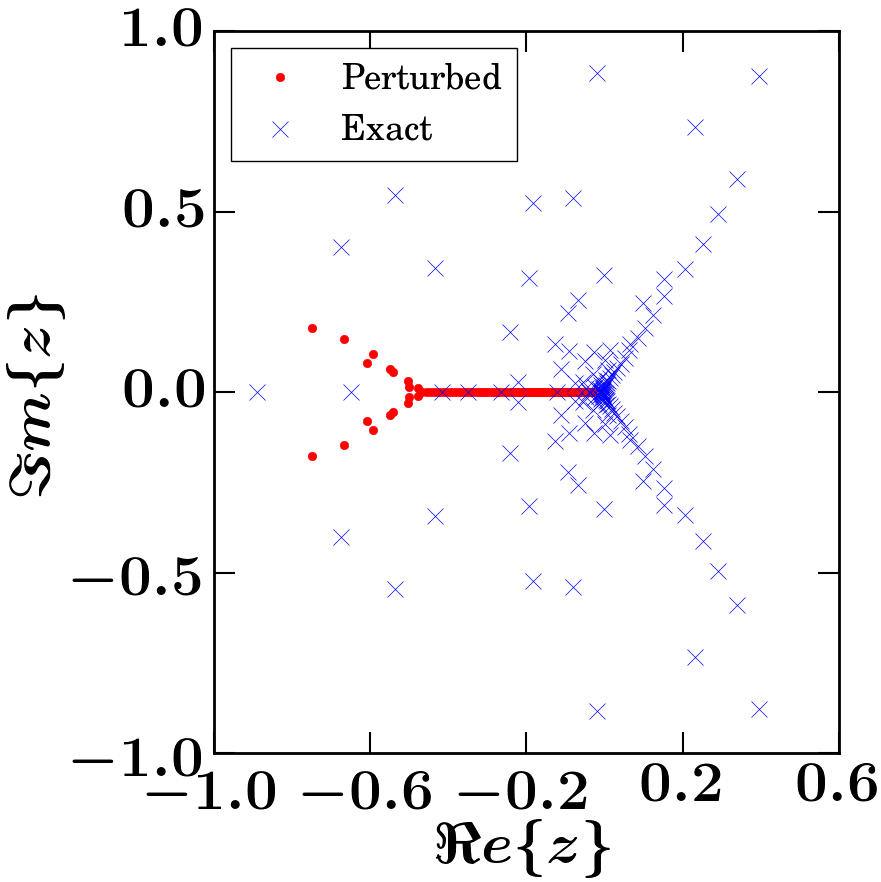}}
    \subfigure[$~T= 0.5 Tc$]{
    \includegraphics[width=0.23\textwidth]{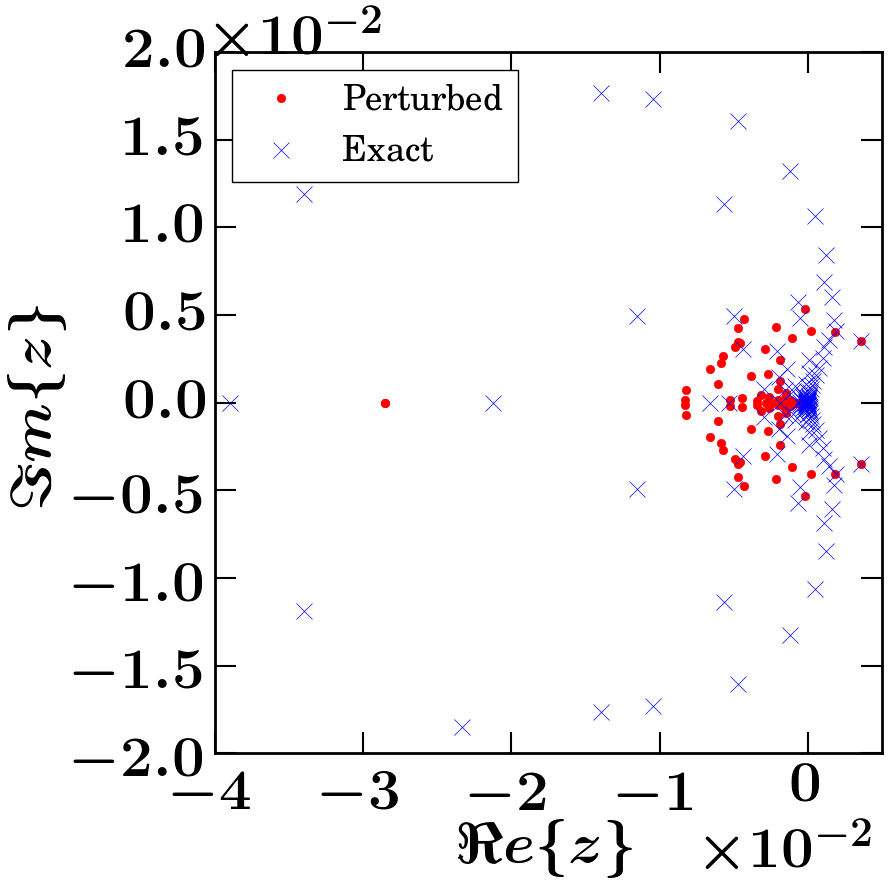}}
    \caption{Comparison between the Yang-lee zeros of the partition function polynomial with exact coefficients (red dots) and the zeros of the partition function with perturbed coefficients (blue crosses). The coefficients are perturbed randomly so that the relative error for all the coefficients is of the order of $10^{-6}$. We see that a drastically different picture of roots appears if the coefficients are not accurate.  Therefore, numerical methods (such as Monte Carlo simulations) to approximate the coefficients, and subsequently the zeros, will be ineffective. The size of the lattice is $16\times16$ with open boundary conditions and $T_c=\frac{2}{\ln{(1+\sqrt{2})}}\frac{J}{k_B}$, where $k_B$ is Boltzmann's constant.}
    \label{fig:PertRoots}
\end{figure}

Bhanot's algorithm still has memory and computation time limitations. This results in difficulty in studying larger system sizes. Moreover, for the AFM case, the roots of the partition function are highly sensitive to any perturbation in the coefficients of the partition function as seen in Fig.~\ref{fig:PertRoots}. Hence, any approximation method, like Monte Carlo simulations for example, results in a drastically different picture of the roots. We, therefore, study the more tractable mean-field model with infinite-ranged coupling between all sites in the next section. 

\section{The Mean-field polynomials}
\label{sec:MFPoly}
We start with the FM case for which we consider $N_s$ spins with all-to-all coupling. The exact Hamiltonian is given by,
\begin{equation}
    \mathcal{H}_{\text{FM}}=\frac{J}{2N_s}\sum_{i\neq j}(1-\sigma_i\sigma_j)-h\sum_{i}(1+\sigma_i),
    \label{MFHamiltnoianFM}
\end{equation}
with $J>0$. The partition function is given by   
\begin{equation}
    \mathcal{Z}_{\text{FM}}=e^{-\frac{1}{2}kN_s }\sum_{M=0}^{N_s}\binom{N_s}{M}z^{M}e^{\frac{1}{2} k N_s \left(2\frac{M}{N_s}-1\right)^2},
    \label{FMMeanPoly}
\end{equation}
where $M=\frac{1}{2}\sum_{i}(1+\sigma_i)$, and $k=\beta J$. Katsura showed that the roots of the polynomial in Eq.~(\ref{FMMeanPoly}) in the complex fugacity plane behave similarly to the FM Yang-Lee zeros of the Ising model~\cite{katsura1954phase}. 

For the AFM interaction, we assume that the system is defined on a bipartite lattice with sub-lattices $A$ and $B$ each containing $\frac{1}{2} N_s$ spins, where a spin on sub-lattice $A$ interacts with every spin on sub-lattice $B$. The Hamiltonian for the AFM case is 
\begin{equation}
    \mathcal{H}_{\text{AFM}}=\frac{2J}{N_s}\sum_{i\in A}\sum_{j\in B}(1-\sigma_i\sigma_j)-h\sum_{i\in A \cup B}(1+\sigma_i).
    \label{MFHamiltnoianAFM}
\end{equation}   

\begin{figure}[t!]
    \centering
    \subfigure[$~T=8 Tc$]{
    \includegraphics[width=0.23\textwidth]{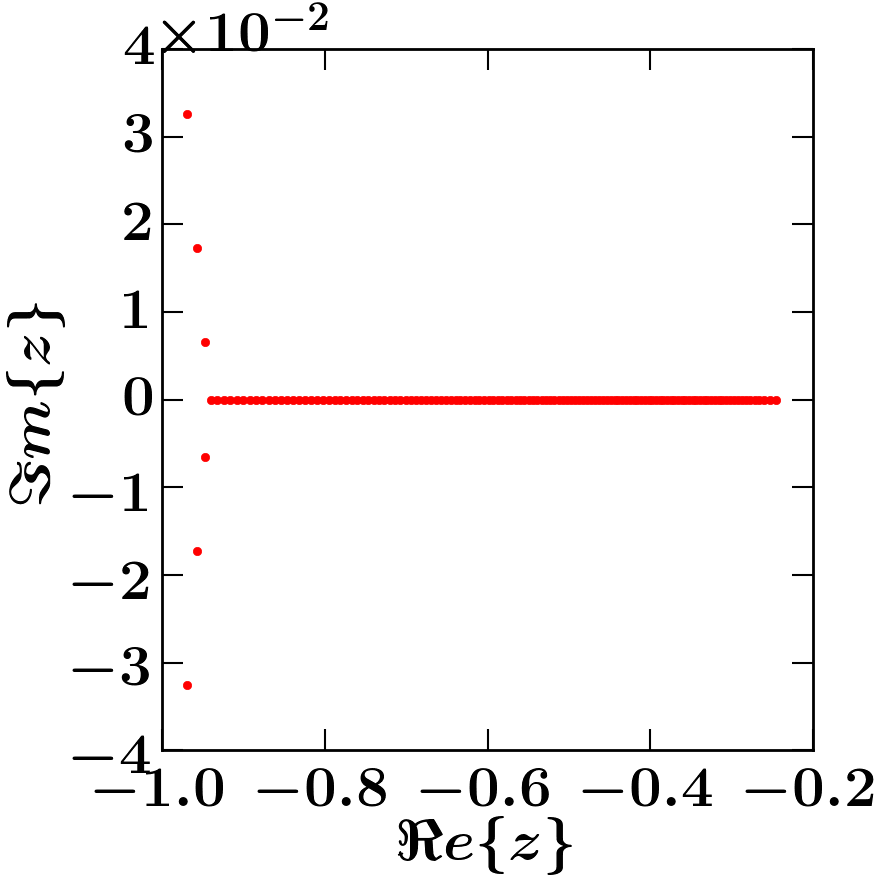}}
    \subfigure[$~T= 2 Tc$]{
    \includegraphics[width=0.23\textwidth]{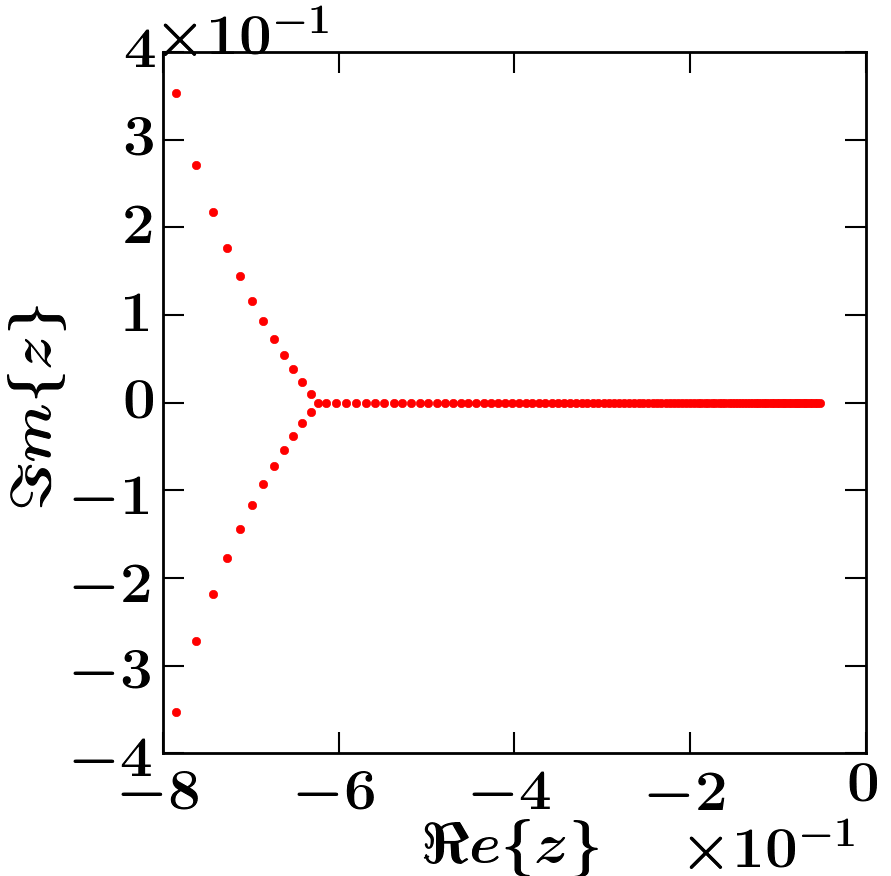}}
    \subfigure[$~T=Tc$]{
    \includegraphics[width=0.23\textwidth]{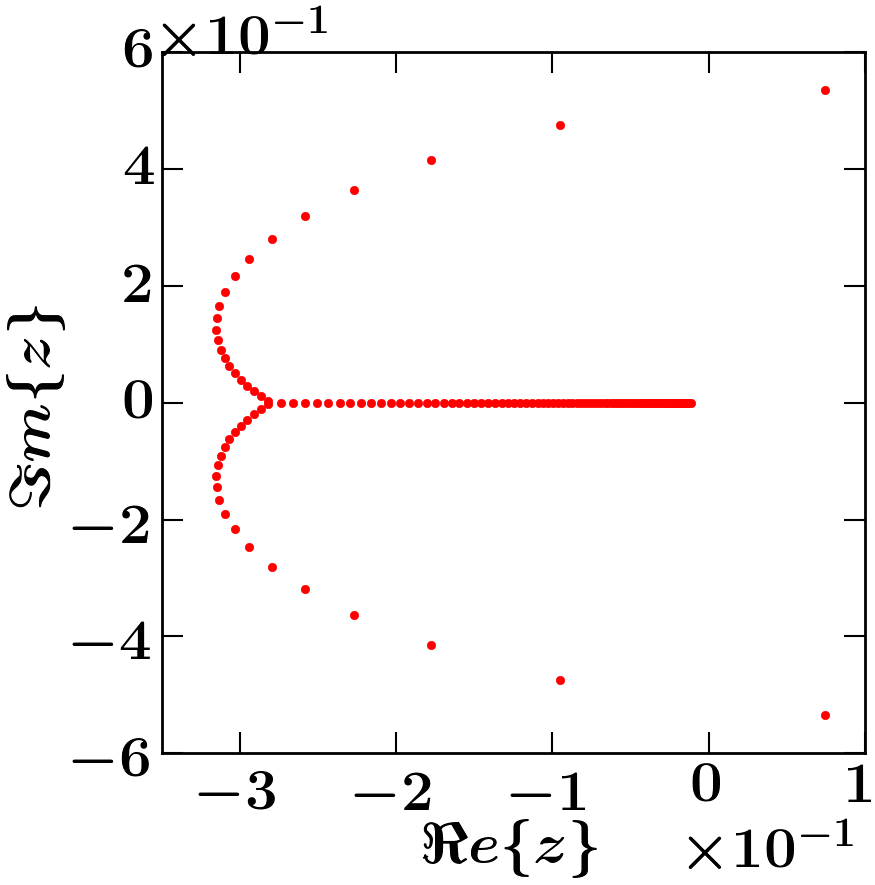}}
     \subfigure[$~T=0.25Tc$]{
    \includegraphics[width=0.23\textwidth]{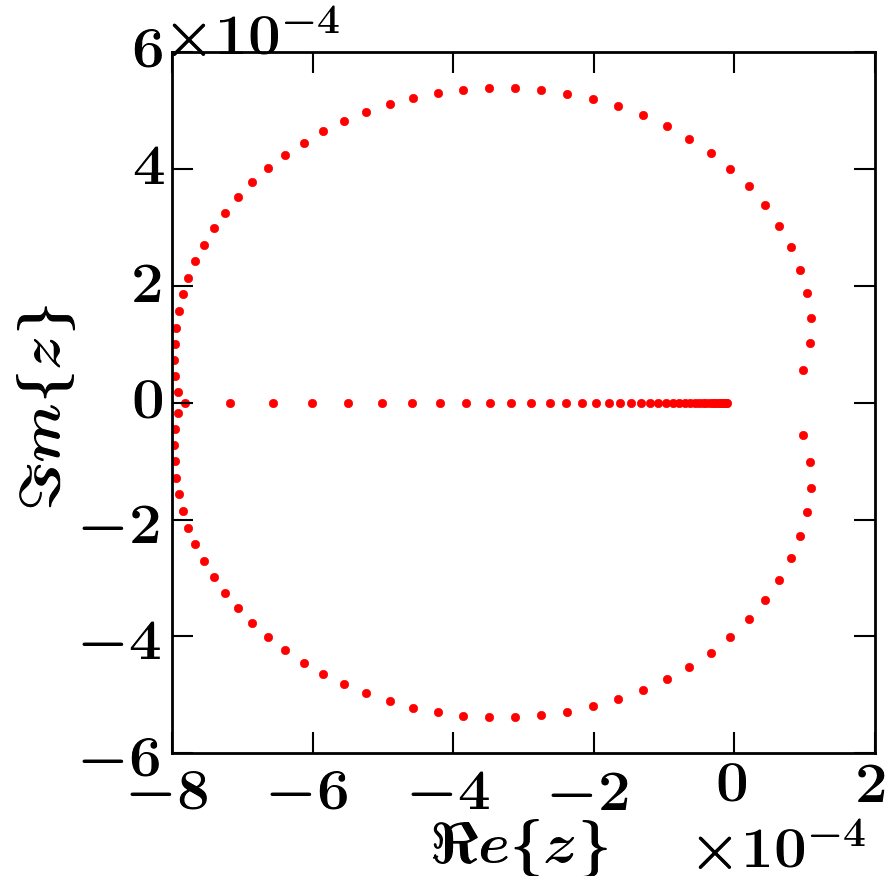}}
    \caption{Yang-Lee zeros in the complex $z$-plane of the AFM mean-field polynomial in Eq.~(\ref{AFMMeanPoly}) with $N_s=16^2$. (a)-(d) show different temperatures. As the temperature decreases, the roots fly off the negative real axis and enter the positive real half-plane. The roots are much more well-behaved than the square lattice Ising model. $T_c=\gamma \frac{J}{k_B} \implies k_c=1$, where $k_B$ is Boltzmann's constant.}
    \label{fig:MFExactRoots}
\end{figure}

The partition function is 
\begin{align}
     \mathcal{Z}_{\text{AFM}}= e^{\frac{1}{2}kN_s}&\sum_{M_a=0}^{N_s/2}\sum_{M_b=0}^{N_s/2} \binom{\frac{1}{2}N_s}{M_a}\binom{\frac{1}{2}N_s}{M_b}z^{M_a+M_b} \notag\\
     &\times e^{-\frac{1}{2} k N_s (4 M_a/N_s-1)(4 M_b/N_s-1)}, 
    \label{AFMMeanPoly}
\end{align}
 where $M_a=\frac{1}{2}\sum_{i\in A}(1+\sigma_i)$, $M_b=\frac{1}{2}\sum_{i \in B} (1+\sigma_i)$, and $k\equiv \beta |J|$. Note that $k>0$ for both FM and AFM cases.  
The behavior of the AFM mean-field (MF) polynomial roots is shown in Fig.~\ref{fig:MFExactRoots} at different temperatures. The roots of the polynomial behave similarly to the AFM Ising model roots that are shown in Fig.~\ref{fig:IsingRoots} in the sense that all roots are real and negative at high temperatures. Upon lowering the temperature, complex roots with $\Re\{z\}<0$ start appearing. Further, reducing the temperature, the roots enter the positive half-plane similar to the AFM Ising model.  Given this similarity with the AFM Ising model and the simplicity of the mean-field models we, therefore, look at partition function polynomials of the mean-field models at high temperatures and its zeros in the thermodynamic limit.

\begin{figure}[t!]
    \centering
    \subfigure[$~\Re\{h\}=-1.71$]{    \includegraphics[width=0.45\textwidth]{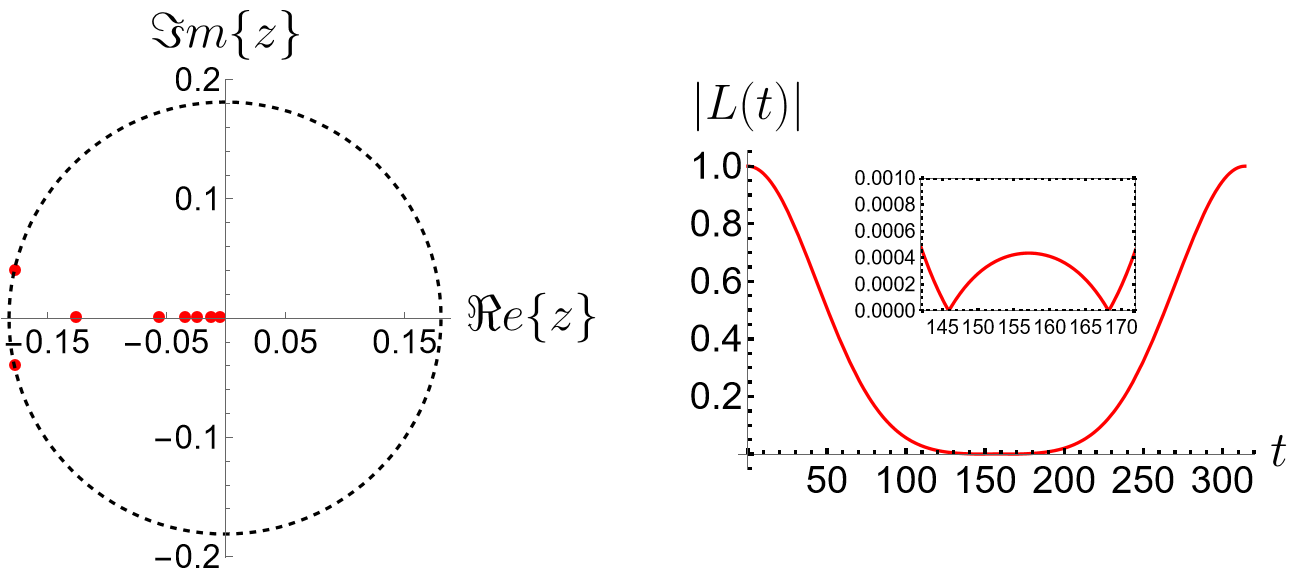}}
    \subfigure[$~\Re\{h\}=-2.08$]{
    \includegraphics[width=0.45\textwidth]{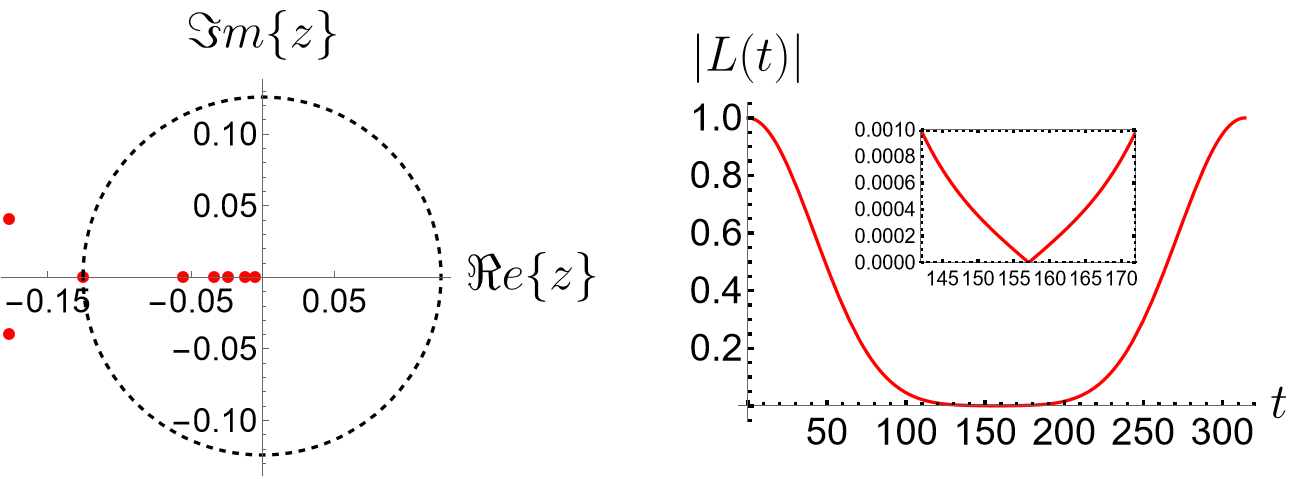}}
    \caption{Correspondence between Yang-Lee zeros and times $t$ at which the coherence $|L(t)|$ vanishes. The results are for the AFM Ising model on a $4\times 4$ lattice at inverse temperature $\beta=0.5$, $\lambda=0.01$, and at two different real magnetic fields indicated by the sub-captions. For the real magnetic field in (a), there are two roots in the complex $z$-plane as shown on the top-right panel, while on the top-left panel, the coherence vanishes at two times corresponding to those roots. Similarly, for (b), there's only 1 root in the bottom-right panel and its corresponding time of vanishing $|L(t)|$ in the bottom-left panel.}
    \label{fig:IsingL(t)}
\end{figure}

\begin{figure}[t!]
    \centering
    \subfigure[$~\Re\{h\}=-0.505$]{    \includegraphics[width=0.45\textwidth]{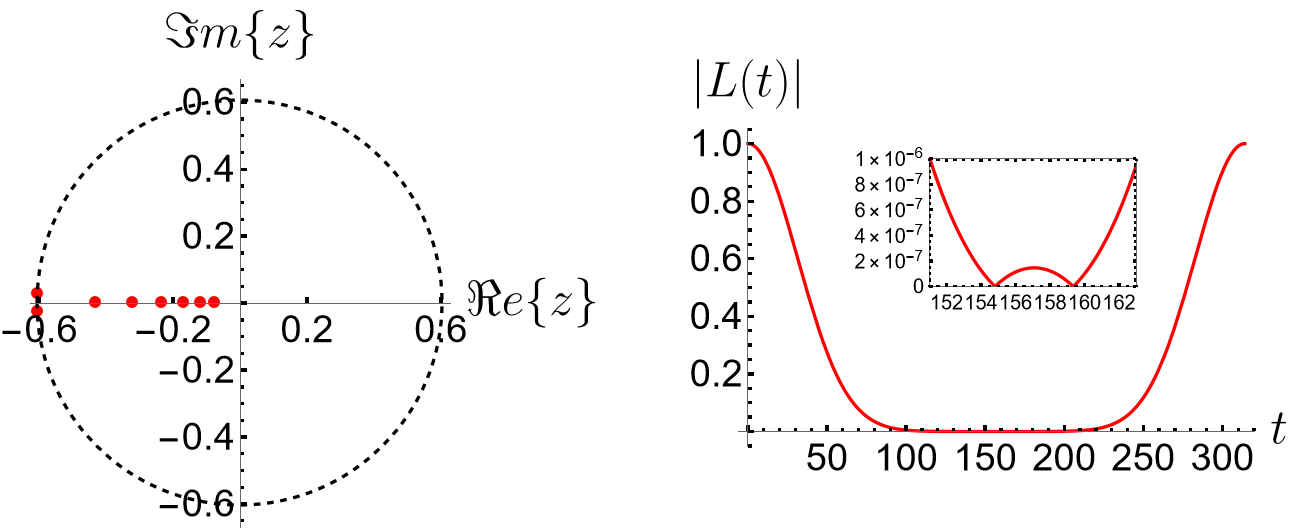}}
    \subfigure[$~\Re\{h\}=-0.841$]{
    \includegraphics[width=0.45\textwidth]{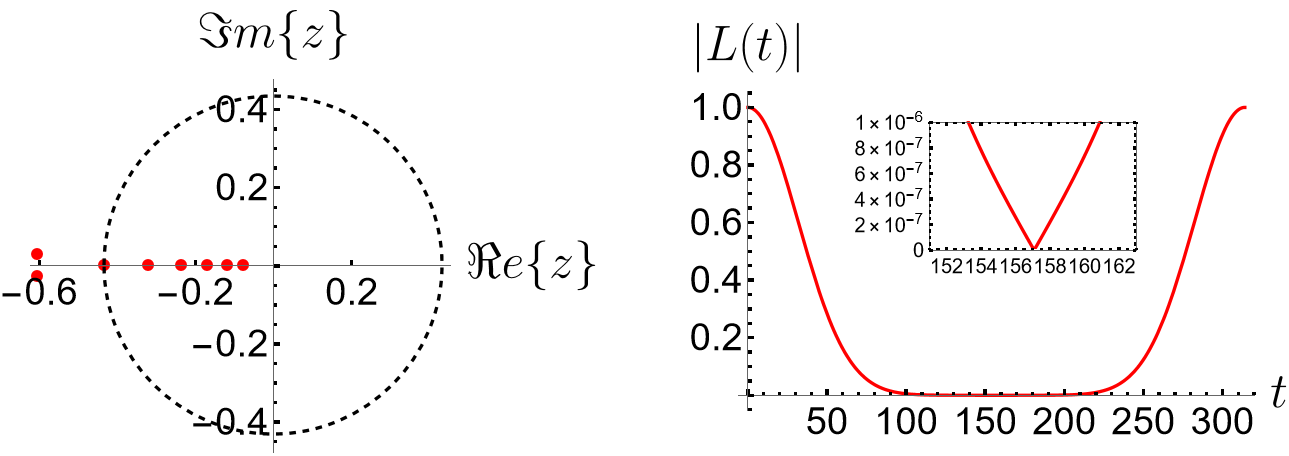}}
    \caption{Correspondence between Yang-Lee zeros and times $t$ at which the coherence $|L(t)|$ vanishes. The results are for the AFM mean-field polynomial for  $N_s= 16$ at inverse temperature $\beta=0.5$, $\lambda=0.01$, and at two different real magnetic fields indicated by the sub-captions. For the real magnetic field in (a), there are two roots in the complex $z$-plane as shown on the top-right panel, while on the top-left panel, the coherence vanishes at two times corresponding to those roots. Similarly, for (b), there's only 1 root in the bottom-right panel and its corresponding time of vanishing $|L(t)|$ in the bottom-left panel.}
    \label{fig:MFL(t)}
\end{figure}

Another way to visualize Yang-Lee zeros was introduced by Wei and Liu who showed a one-to-one correspondence between the Yang-Lee zeros and the times at which the coherence of a probe spin coupled to a many-body system vanishes~\cite{wei2012lee}. They showed that the probe spin coherence as a function of time $t$ is given by 
\begin{align}
     L(t)\propto \frac{\prod_{j=1}^{N_s} (e^{2\beta(\Re\{h\}+2i\lambda t/\beta)}-z_j)}{\prod_{j=1}^{N_s}(e^{2\beta(\Re\{h\})}-z_j)}
     \label{L(t)},
\end{align}
where $z_j$ are the Yang-Lee zeros of the partition function of the corresponding many-body system. Therefore, at a given real magnetic field $\Re\{h\}$, the times at which $|L(t)|$ vanishes correspond to Yang-Lee zeros. Wei and Liu studied the FM Ising model where all the roots lie on the unit circle in $z$-plane $(\Re\{h\}=0)$. We show the correspondence between Yang-Lee zeros and $|L(t)|$ for the AFM Ising model and AFM mean-field polynomial in Figs. \ref{fig:IsingL(t)} and \ref{fig:MFL(t)}, respectively.

In the next subsection, we show that the logarithm of Yang-Lee zeros for the AFM mean-field model scale as $\sim \sqrt{k}$ at high temperatures similar to the roots of the exact nearest-neighbor AFM Ising model. We obtain this result by two different methods. The former is by applying the procedure in subsection \ref{subsec:HighTemp}, and the latter is by showing that the AFM mean-field polynomial at high temperatures is a linear combination of Hermite polynomials in the variable $\xi$. This linear combination leads to roots, $\xi_j$, to scale as powers of $\sqrt{k}$ at high temperatures.

\subsection{AFM MF Polynomial at High Temperatures}
The goal of this subsection is to show that the logarithm of Yang-Lee zeros of the AFM mean-field polynomial defined in Eq.~(\ref{AFMMeanPoly}) scale as $\sqrt{k}$ at high temperature. The first method is
to apply the procedure in Sec. \ref{subsec:HighTemp}. We compare the expansion in Eq.~(\ref{PolyExpand}) to the high-temperature expansion of the AFM MF polynomial. We expand the exponential in Eq.~(\ref{AFMMeanPoly}) for high temperature ($\frac{kN_s}{2}\sim0$) to get
\begin{align}
   &\mathcal{Z}_{\text{AFM}}= \sum_{M_a=0}^{N_s/2}\sum_{M_b=0}^{N_s/2} \binom{\frac{1}{2}N_s}{M_a}\binom{\frac{1}{2}N_s}{M_b}z^{M_a+M_b}
   \Bigg[1\notag\\
   &+\left(1-\left(4\frac{M_a}{N_s}-1\right)\left(4\frac{M_b}{N_s}-1\right)\right)\frac{kN_s}{2}\notag\\
   &+\frac{1}{2}\left(4\frac{M_a}{N_s}-1\right)^2\left(4\frac{M_b}{N_s}-1\right)^2\left(\frac{kN_s}{2}\right)^2+\dots\Bigg]. 
   \label{AFMHighTExpand}
\end{align}
The summations can be carried out separately for each order of $\Tilde{k}$ using binomial expansions. This results in the high-temperature expansion for the AFM mean-field polynomial as 
\begin{align}
    \frac{1}{N_s}\ln{\mathcal{Z}_{\text{AFM}}}&=\ln{(1+z)}+\frac{2z}{(1+z)^2}k\notag\\
    &+2\left(\frac{(z-2z^2+z^3)+\frac{4z^2}{N_s}}{(1+z)^4}\right)k^2+\mathcal{O}(k^3).
    \label{HighTExpandMF}
\end{align}
Since all the roots $\xi_j$ approach 0 at high temperatures, we can assume the same expansion as in Eq.~(\ref{Xiofk}). Hence, by plugging this expansion into Eq.~(\ref{PolyExpand}) then comparing the result with Eq.~(\ref{HighTExpandMF}), we reach the same conclusion for the MF polynomials that the logarithm of Yang-Lee zeros scale as $\sqrt{k}$ at high temperatures similar to the nearest neighbor Ising model as in Eq.~(\ref{Xibehave}). We can also derive similar sum rules to Eqs.~(\ref{Sum-Rule1},\ref{Sum-Rule2}) by comparing the coefficients of $k$ in Eq.~(\ref{HighTExpandMF}) with Eq.~(\ref{PolyExpandk}). This gives the sum rules for the MF model 
\begin{equation}
    \mathcal{A}=1, \quad \mathcal{B}=\frac{1}{2N_s}\left(1+\frac{1}{2}N_s\right), \quad \mathcal{C}=\frac{2}{N_s}\left(-1+N_s\right).
    \label{Sum-Rule12MF}
\end{equation}

The second method to obtain the $\sqrt{k}$ result is to write the AFM mean-field polynomial defined in Eq.~(\ref{AFMMeanPoly}) as a linear combination of Hermite polynomials by expressing it in an integral form and then approximating the integrand at high temperatures.  This approximation is given by  
\begin{equation}
    \mathcal{Z}_{\text{AFM}}\approx\left(-\alpha e^{-\alpha \tilde{\xi}}\right)^{N_s}\sum_{p=0}^{N_s/2}(-1)^p\binom{\frac{N_s}{2}}{p}\frac{(2p)!}{p!}H_{N_s-2p}(\tilde{\xi}).
    \label{AFMHermite}
\end{equation}
where $\tilde{\xi}=\sqrt{\frac{2k}{N_s}}\xi$, with $\xi$ defined in Eq.~(\ref{DefineXi}). The detailed derivation of Eq.~(\ref{AFMHermite}) is given in appendix \ref{Appx:HermiteDerivation}. The roots $\tilde{\xi}_j$ are related to the roots $\xi_j$ defined in by 
\begin{equation}
    \xi_j=\alpha \Tilde{\xi}_j=\sqrt{k}\sqrt{\frac{2}{N_s}}\tilde{\xi}_j.
    \label{HighTempRootDependence}
\end{equation}
Since the roots $\tilde{\xi}_j$ are independent of temperature, the roots of the AFM mean-field polynomial have $\sqrt{k}$ dependence at high temperature. 

 The free energy associated with the partition functions that are described by the mean-field polynomials could be written down explicitly and used to determine the zeros of these polynomials. In the next section, we show that the density of the zeros in the $z$-plane is related to the free energy of the model via a simple relation. We subsequently use this relation to numerically determine the Yang-lee zeros of the mean-field polynomials in the thermodynamic limit.
\subsection{Density of roots in the thermodynamic limit} 
\label{subsec:free_en_tech}
The real part of the free energy per site of the partition function in terms of the Yang-Lee zeros is given by 
\begin{equation}
    \beta f(z) = \beta \Re\{\Tilde{f}(z)\}=\frac{1}{N_s}\sum_{j=1}^{N_s}\ln|z-z_j|+c,
    \label{FiniteReFE}
\end{equation}
where $c$ is a constant. In analogy with basic electrostatics, the real part of the free energy can be interpreted as the potential due to $N_s$ point charges located at  $z_j$  on a two-dimensional plane. Therefore, if the free energy per site is known, the location of the roots would be revealed by taking the Laplacian of its real part. In fact, in the thermodynamic limit, the Laplacian of the real part of the free energy would directly give us the density of the roots, $\rho(z)$. We can therefore write, 
\begin{equation}
    \rho(z)= -\frac{1}{2\pi}\nabla^2(\beta f(z)).
    \label{PoissonEq}
\end{equation}
A similar approach of calculating the density of Lee-Yang zeros was explained in Bena, et all~\cite{bena2005statistical}. Hemmer and Hauge~\cite{hemmer1964yang} used this technique to study the Yang-Lee zeros of a van der Waals gas. We use Eq.~(\ref{PoissonEq}) to numerically determine the Yang-Lee zeros for the AFM and FM mean-field models in the thermodynamic limit. For these models, the free energy per site can be computed by using the saddle point approximation of the corresponding partition functions. To that end, we first note that for large $N_s$, we can use  Stirling's formula to write the FM  and AFM polynomial as $\mathcal{Z}_{\text{FM}}=\sum_{M}e^{-N_s \beta \Tilde{f}_{\text{FM}}}$ and  $\mathcal{Z}_{\text{AFM}}=\sum_{M_a,M_b}e^{-\frac{1}{2}N_s \beta \Tilde{f}_{\text{AFM}}}$, respectively. $\tilde{f}_{\text{FM}}$ and $\tilde{f}_{\text{AFM}}$  are given by

\begin{align}
    &\beta \Tilde{f}_{\text{FM}}(m)=-\frac{1}{2} k m^2- \frac{1}{2} \ln{(z)} m- s(m)+\frac{1}{2}k,\label{FMfreeEnergy}\\
    &\beta \Tilde{f}_{\text{AFM}}(m_a,m_b)=k m_a m_b-  \frac{1}{2} \ln{z}( m_a + m_b)\notag\\
    &\hspace{0.15\textwidth}- \frac{1}{2} s(m_a) - \frac{1}{2} s(m_b)-k, \label{AFMfreeEnergy}
\end{align}
where
\begin{align}
    &s(m)=-\left[\frac{1+m}{2}\ln{\left(\frac{1+m}{2}\right)}+\frac{1-m}{2}\ln{\left(\frac{1-m}{2}\right)}\right],
    \label{EntropyAFM}
\end{align}
 $m=2\frac{M}{N_s}-1$, and $m_{a/b}=2\frac{M_{a/b}}{N_s}-1$ are the magnetizations per site.

For the saddle point approximation, we first consider the FM case. We use Eq.~(\ref{FMfreeEnergy}) to write the partition function for large $N_s$ as a sum of exponential terms using the variable $m_i=2i/N_s-1$ with $\Delta m = m_{i+1}-m_{i}=2/N_s$ as $\mathcal{Z}_{\text{FM}}=\frac{N_s}{2}\sum_{i=0}^{N_s}e^{-N_s \beta \Tilde{f}_{\text{FM}}(m_i)}\Delta m$. This sum could be approximated as an integral over a real continuous variable $-1 \leq m \leq 1$. If we allow $m$ to be complex, the contour of integration could be deformed into another contour $C$, which has the same endpoints and passes through saddle points $m^*$ satisfying $\frac{\partial \Tilde{f}}{\partial m}\big|_{m^*}=0$ in the complex $m$-plane. The partition function could then be approximated using saddle point approximation over the new contour as 
\begin{align}
   &\frac{N_s}{2}\sum_{i=0}^{N_s}e^{-N_s \beta \Tilde{f}_{\text{FM}}(m_i)}\Delta m \approx \int_C e^{-N_s \beta \Tilde{f}(m)} \mathrm{d}m\notag\\
   & \sim e^{-N_s \beta \Tilde{f}_{\text{FM}}(m_1^*)}+e^{-N_s \beta \Tilde{f}_{\text{FM}}(m_2^*)}+\dots \ .
    \label{PFSaddles}
\end{align}
Note that the new contour $C$ has to remain within the branch cuts of $\Tilde{f}_{\text{FM}}(m)$ closest to the real axis so that the value of the integral does not change. Therefore, only saddle points within those branch cuts should be considered. As $N_s \rightarrow \infty$, the saddle point $m^*$ that minimizes the real part of $\Tilde{f}(m)$ will have the most contribution and dominate the other saddles. Therefore, $\Tilde{f}_{\text{FM}}(m^*)$ with the minimum real part is the free energy per site of the system in the thermodynamic limit. The saddle points are given by the mean-field theory (MFT) self-consistency equation 
 \begin{equation}
      m^*= \tanh{(km^*+\beta h)}.   
     \label{FMsaddleEq}
 \end{equation} 
A similar calculation for the AFM case to derive the AFM version of Eq.~(\ref{PFSaddles}) could be done by applying the saddle point approximation for the variables $m_a$ and $m_b$. Hence, we get the saddle points by setting $\frac{\partial \Tilde{f}_{\text{AFM}}}{\partial m_a}=0$, and $\frac{\partial \Tilde{f}_{\text{AFM}}}{\partial m_b}=0$, which lead to the equations 
 \begin{align}
         m_a^*=\tanh(-km_b^*+\beta h),\label{AFMsaddleEqs1}\\
         m_b^*=\tanh(-km_a^*+\beta h),
     \label{AFMsaddleEqs2}
 \end{align}
 respectively.

\subsection{Numerical results}
 In the numerical implementation to find the roots using free energy, we first choose a region in either $z$-plane or $\xi$-plane that includes the roots. We then choose a fine grid for the region in either plane and find the free energy for each point on the grid. Once the free energy is known for each point on the grid, the density of the roots is then calculated by taking the Laplacian of the free energy numerically.\\

\noindent
\textit{Numerical results for FM:}\label{subsec:Iteration} To determine the free energy per site in the thermodynamic limit, the problem boils down to finding the saddle point $m^*$ that minimizes the real part of free energy. %%We show two techniques to find saddle points of Eqs. (\ref{FMsaddleEq}, \ref{AFMsaddleEqs}): iterative technique, and a numerical algorithm called simplicial homology algorithm. We view the former in this subsection. 
One of the simplest ways is to consider the iterations of the saddle point equation, Eq.~(\ref{FMsaddleEq}). 
%Iterations of the saddle point Eqs. (\ref{FMsaddleEq}, \ref{AFMsaddleEqs}) for FM and AFM are of the form
So, we consider the following iteration scheme for the FM case
\begin{equation}
    \begin{split}
        &m_{i+1}=g_{\text{FM}}(m_{i})=\tanh{(km_i+\beta h)},\\
    \end{split}
    \label{IterEqFM}
\end{equation}
This iteration equation shows a fast convergence to some fixed point, $m^*$, which would therefore be a solution to the saddle point equation. For the FM case, the saddle point that minimizes $\Re\{\Tilde{f}_{\text{FM}}(m)\}$ is always an attractive fixed point of the mapping $g_{\text{FM}}(m)$. Therefore, iteration is enough to find the solution which minimizes the real part of the free energy.

\begin{figure}[t!]
    \centering
    \subfigure[$~k=0.25$]{
    \includegraphics[width=0.23\textwidth]{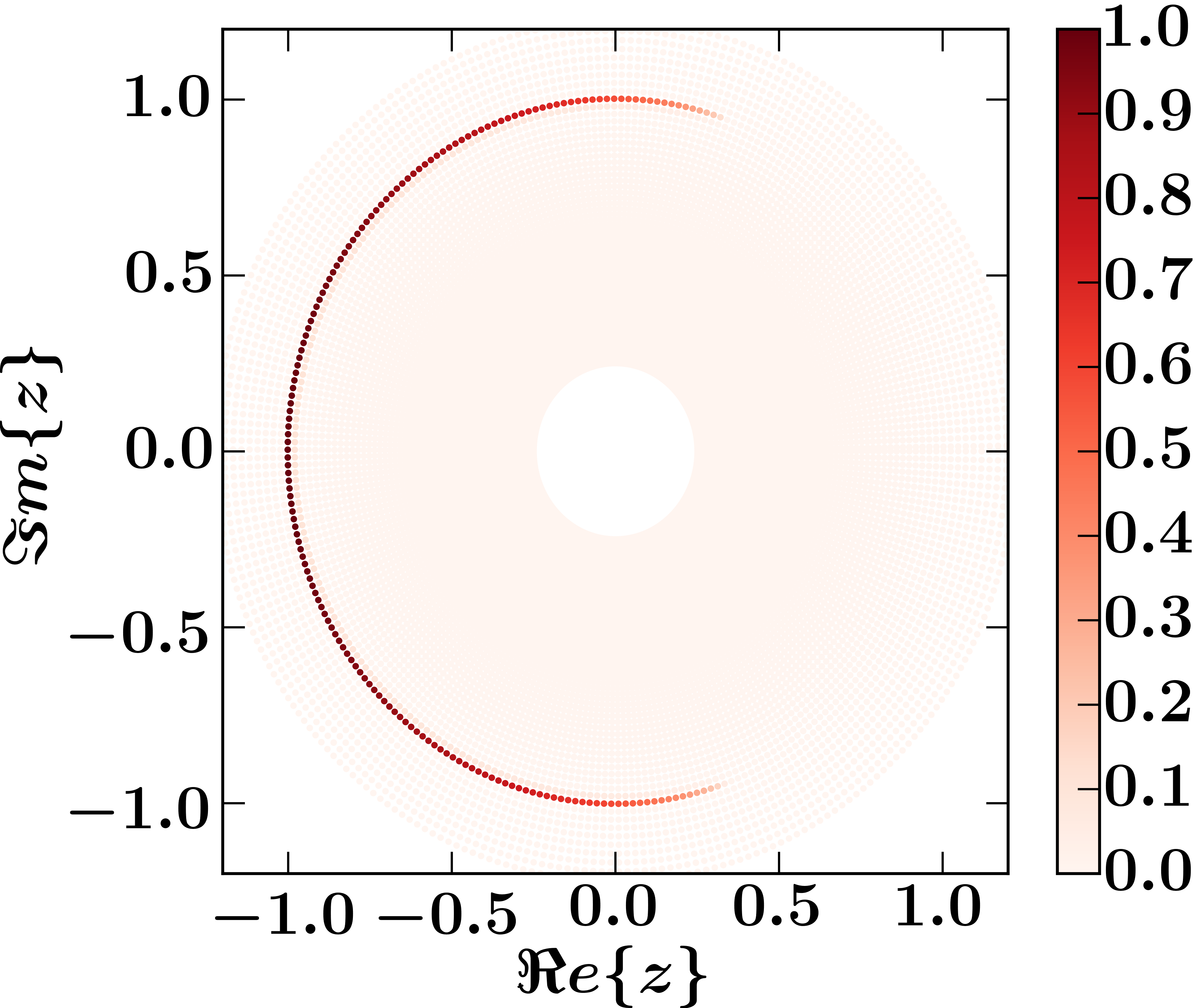}}
    \subfigure[$~k=0.5$]{
    \includegraphics[width=0.23\textwidth]{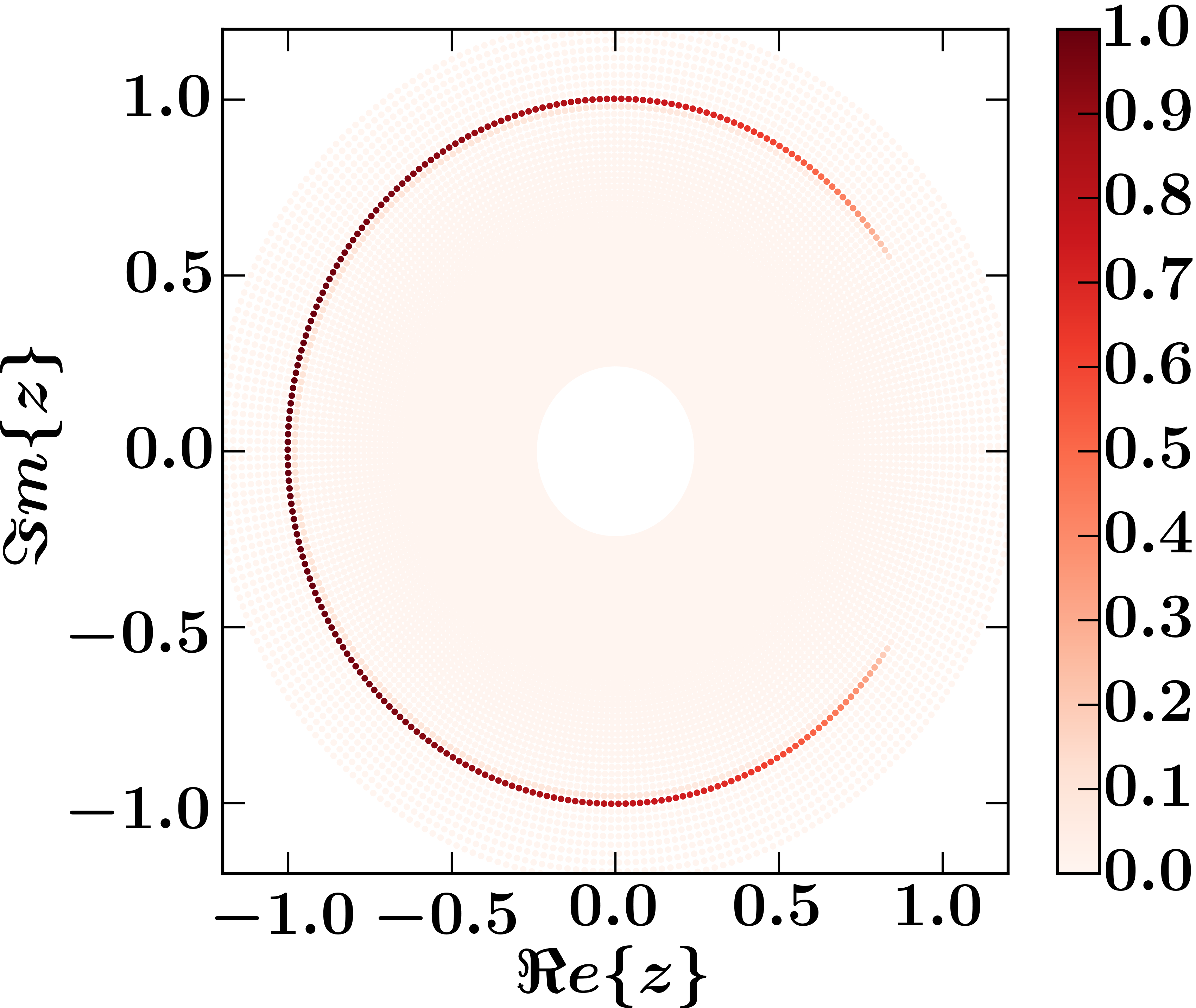}}
    \subfigure[$~k=1$]{
    \includegraphics[width=0.23\textwidth]{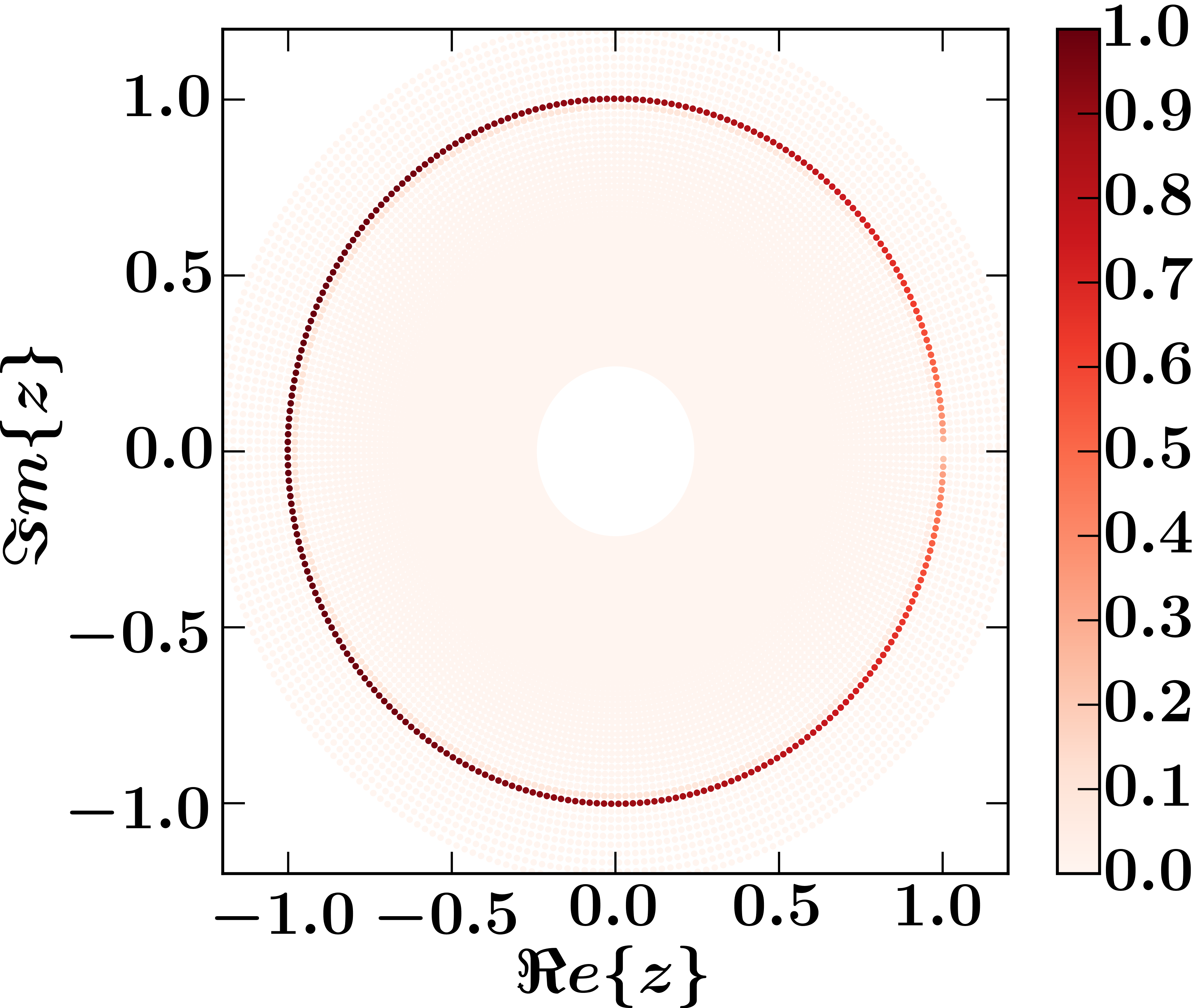}}
    \subfigure[$~k=1.5$]{
    \includegraphics[width=0.23\textwidth]{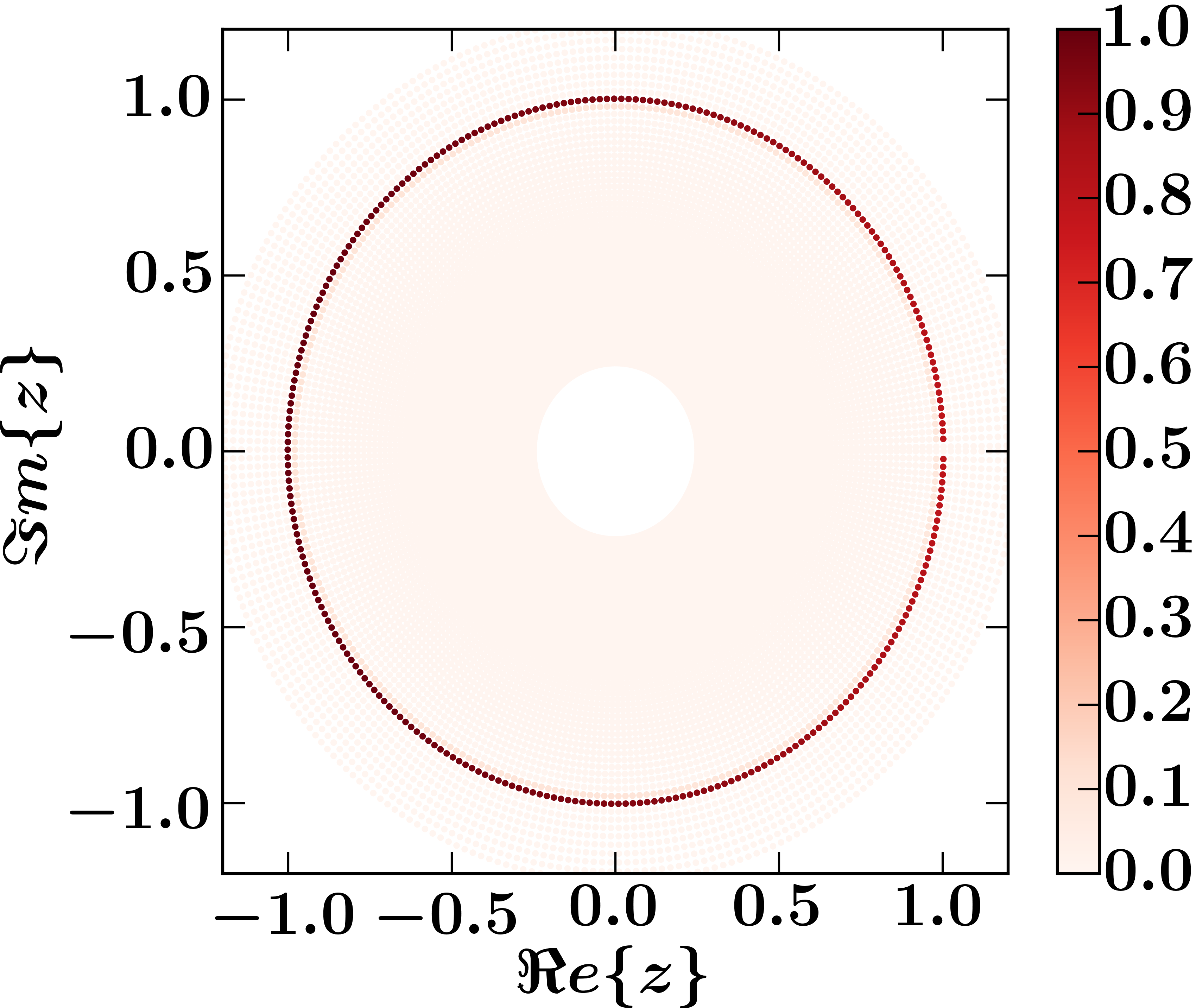}}
    \caption{Contour plot of $\left|\frac{\rho(z)}{\max\{\rho(z)\}}\right|$ in $z$-plane at different values of $k$. The critical temperature at $k_c=1$. The root arc is in agreement with the Yang-Lee theorem as the roots lie on an arc of the unit circle for $k<k_c$ in panels (a) and (b). At the critical temperature, the roots close onto the real axis forming a full circle in panel (c). Further reducing the temperature, the density approaches uniform distribution on the unit circle in panel (d).}
    \label{fig:FM-zroots}
\end{figure}

We take a grid in the $z$-plane around the unit circle because the roots of FM mean-field polynomial lie on the unit circle. We use the iteration scheme in Eq~(\ref{IterEqFM}) to find the free energy for every point in the grid. Numerical calculations of the Laplacian of the free energy give the root density of the FM polynomial in the thermodynamic limit. The results for the density of the roots are shown in Fig.~\ref{fig:FM-zroots}. The results are in agreement with the Yang-Lee theorem since the roots lie on an arc of the unit circle for $k < k_c$ and upon decreasing the temperature, the roots close onto a full circle at $k=k_c$. For $k>k_c$, the roots seem to be approaching a uniform distribution along the unit circle. As mentioned in Sec. \ref{subsec:Iteration}, the iteration always gives the saddle point $m^*$ with the minimum real part of the free energy. 

\begin{figure}[t!]
    \centering
    \includegraphics[width=0.5\textwidth]{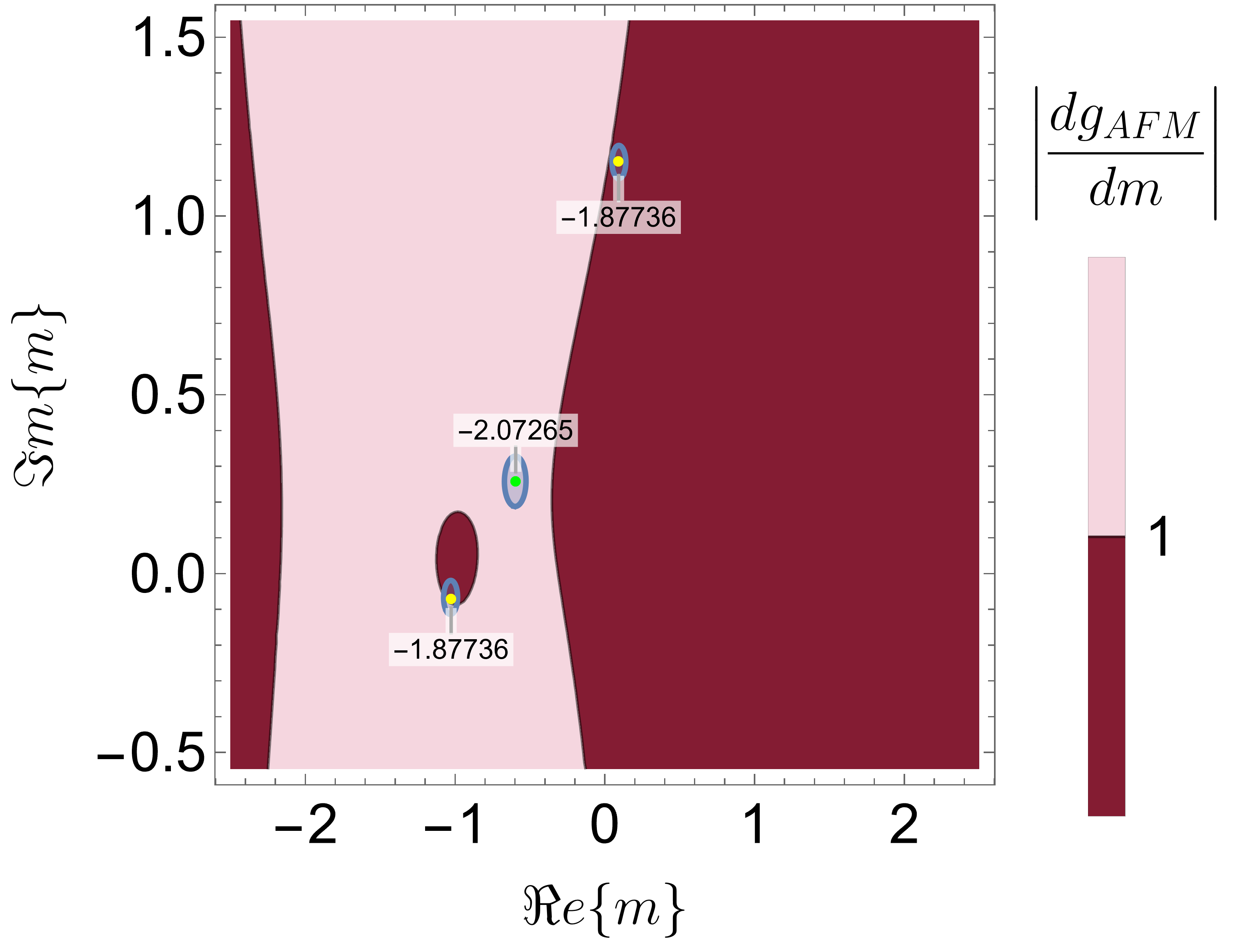}
    \caption{Saddle points of Eq.~(\ref{IterEqAFM}) at $k=1$ and $\beta h = 1.5 - 0.75 i$ in $m$-plane. The yellow saddles are obtained through iteration, whereas the green one is obtained with SGHO. The pink region corresponds to repulsive points $\left|\frac{\mathrm{d}g_{\text{AFM}}}{\mathrm{d}m}\right| > 1$, while the dark red regions are attractive points  $\left|\frac{\mathrm{d}g_{\text{AFM}}}{\mathrm{d}m}\right| > 1$. Each saddle is labeled with the corresponding value of $\Re\{f_{\text{AFM}}(m)\}$. The saddle point with the minimum value is a repulsive fixed point, hence not approachable by iteration.}
    \label{fig:IterFail}
\end{figure}

\vspace{0.3cm}
\noindent
\textit{Numerical results for AFM:} For the AFM case, the iteration scheme could be used as
 \begin{equation}
    \begin{split}
        &m_{i+1}=g_{\text{AFM}}(m_{i})=\tanh{(-k\tanh{(-km_i+\beta h)}+\beta h)},\\
    \end{split}
    \label{IterEqAFM}
\end{equation}
 where $m\equiv m_a$ in the AFM case and $m_b$ is fixed by Eq.~(\ref{AFMsaddleEqs2}). However, such a scheme to find solutions to the saddle point equation does not guarantee that the real part of the free energy would be a global minimum. In fact, it may be possible that the solution that minimizes the real part of the free energy may not be an attractive fixed point at all. An example of this is shown in Fig.~\ref{fig:IterFail} where we show that for a particular value of $k$ and $h$, the fixed point which is repulsive~(green dot) has a smaller real part of the free energy compared to the other two fixed points~(red dots) which are attractive in nature. It is worth mentioning that changing the variable of iteration from the magnetization per site $m_{i}$ to the effective magnetic field $\psi_i=-km_i+\beta h$, and iteration $\psi_{i+1}=-k\tanh{(-k\tanh{(\psi_{i})+\beta h})}+\beta h$, seems to be more effective in finding saddle points with iteration. The quantity $km$ is of the same order as $\beta h$ and this causes underflow precision errors when evaluating the $\tanh$ function. This is fixed by iterating the $\psi_i$ variable.

 \begin{figure}[t!]
    \centering
    \subfigure[$~k=0.5$]{
    \includegraphics[width=0.23\textwidth]{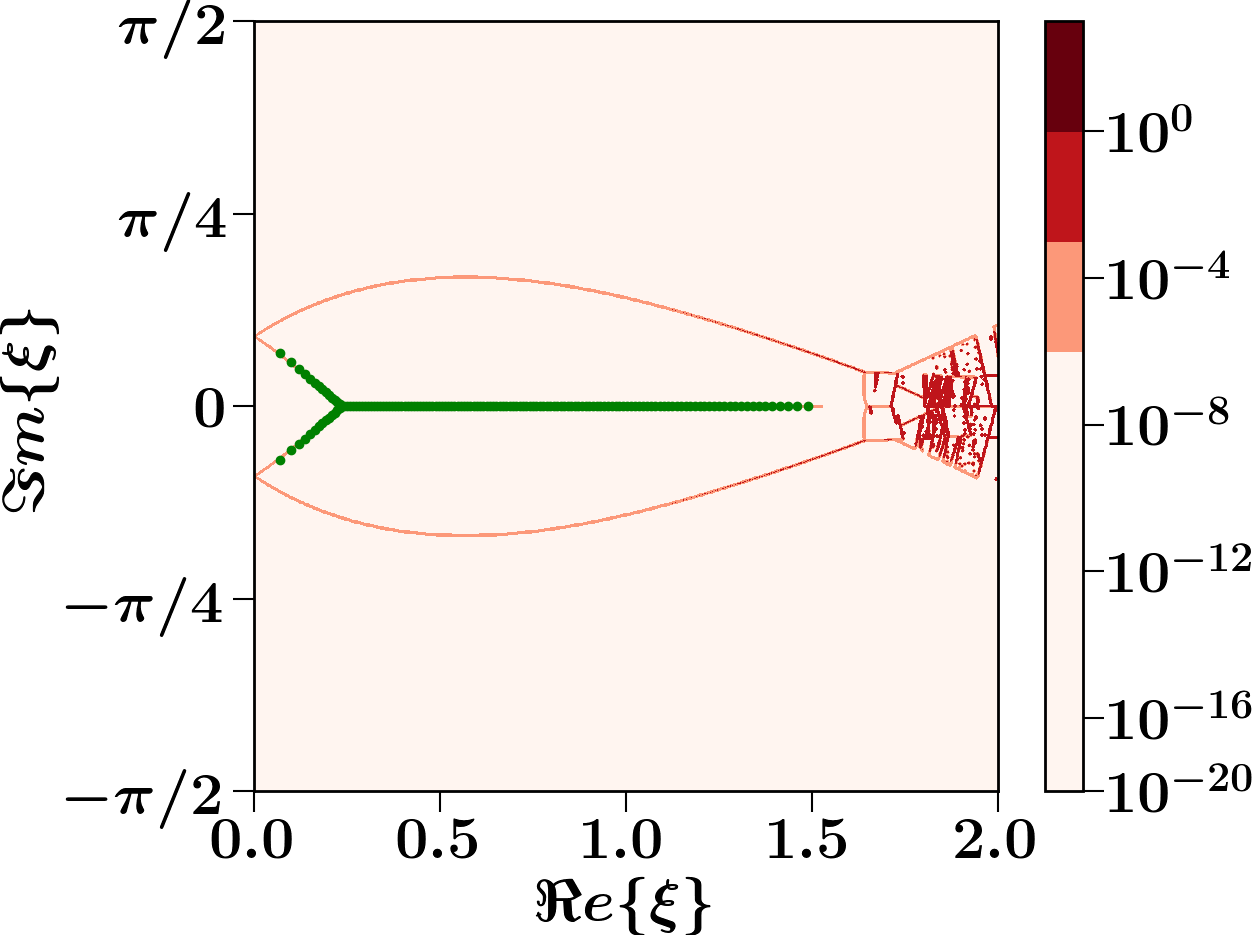}}
    \subfigure[$~k=1$]{
    \includegraphics[width=0.23\textwidth]{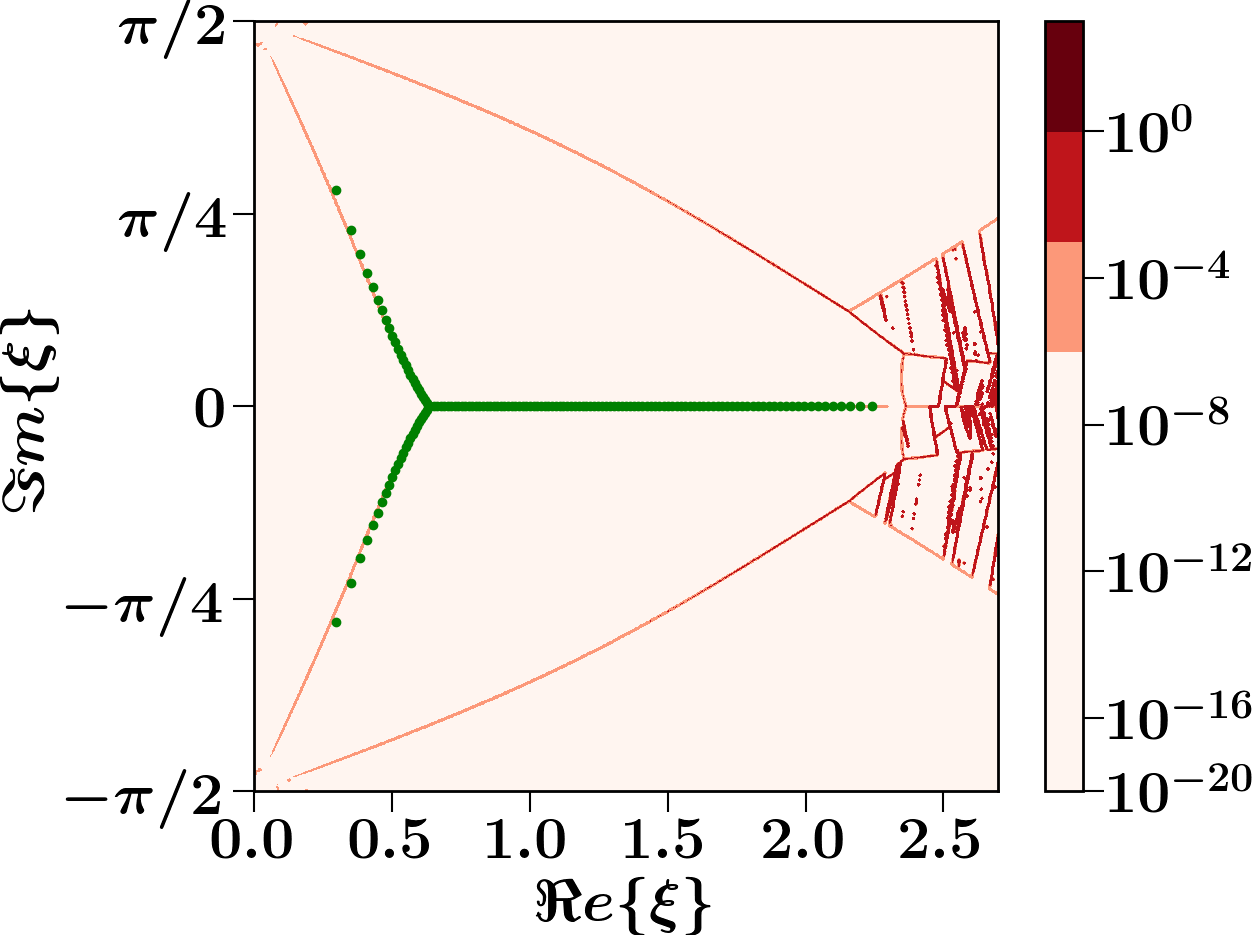}}
    \subfigure[$~k=1.5$]{
    \includegraphics[width=0.23\textwidth]{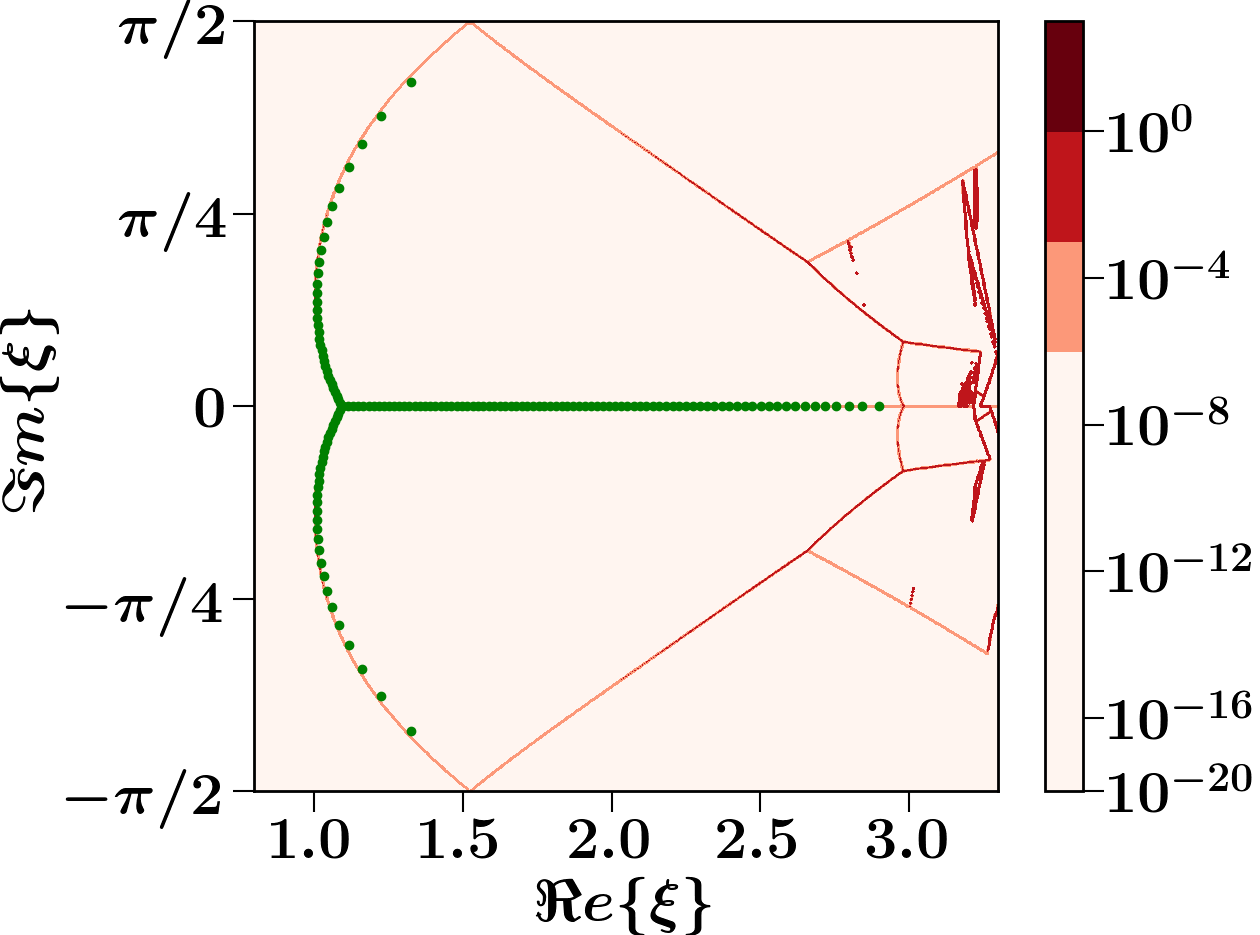}}
    \subfigure[$~k=5$]{
    \includegraphics[width=0.23\textwidth]{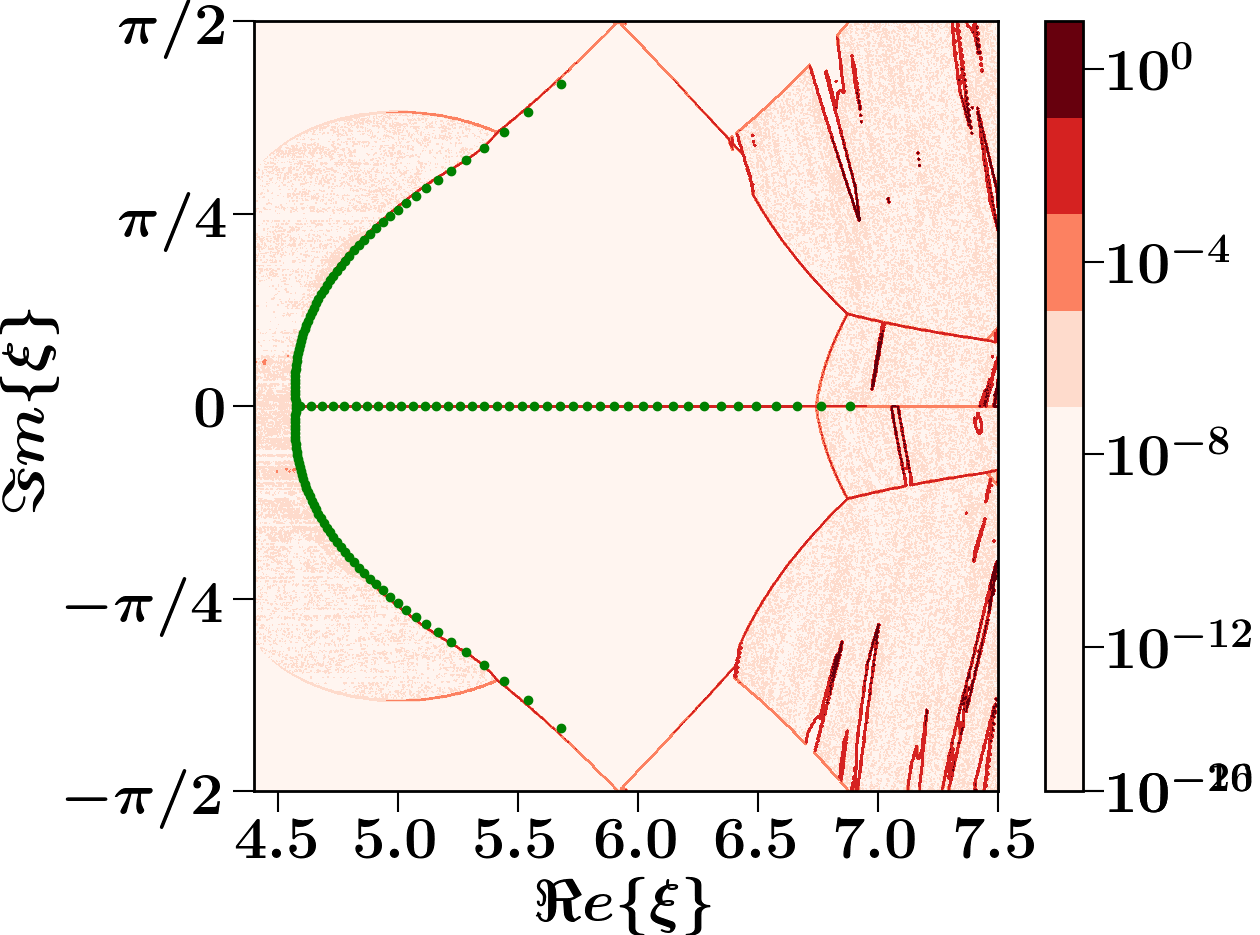}}
    \caption{Contour plot of $\left|\frac{\rho\left(z=-e^{-2\xi}\right)}{\max\{\rho(z)\}}\right|$ in $\xi$-plane at different values of $k$. The exact roots for $N_s=300$ are shown in green. At temperatures greater than the critical temperature (a), the curves do not touch the line $\Im\{\xi\}=\pi/2$. The roots start pinching the line $\Im\{\xi\}=\pi/2$ at $k=k_c=1$ (b). This marks the phase transition point as the pinching point corresponds to a real value of magnetic field $h$. For lower temperatures $k>k_c$ (c) and (d), the pinching point shifts along the line $\Im\{\xi\}=\pi/2$ with an increasing $\Re\{\xi\}$. The critical temperature is $k_c=1$, and the legend bar is given in $\log$ scale.}
    \label{fig:xiroots}
\end{figure}

 To find the solutions to the AFM saddle point equation, we consider the function $|m-g_{\text{AFM}}(m)|$ and look for values of $m$ at which this function has vanishing minima. To that end, we use the so-called \textit{simplicial homology global optimization}~(SHGO) algorithm~\cite{endres2018simplicial}. This algorithm could be used to determine different minima of a given function within a given range by efficiently locating suitable starting points of the search for minima. Once the starting points are known, a local minimization routine can be used to locate a particular minimum. A useful Python implementation of this algorithm can be found in~\cite{endres2016python}.  Using the SHGO algorithm, we find multiple values of $m$ at which the function, $|m-g_{\text{AFM}}(m)|$ has vanishing minima in a given region of the $m$-plane. The $m$ corresponding to these minima are the solutions to the AFM saddle point equations. Out of these saddle points, we choose the one with the lowest real part of $\Tilde{f}_{\text{AFM}}$. Note that the chosen saddle point might not be the global minimum of the real part of $\Tilde{f}_{\text{AFM}}$. 

 We now take a grid in $\xi$-plane instead of $z$-plane because the roots in the $z$-plane cluster around $z=0$ for low temperatures, which leads to precision errors when numerically computing the Laplacian of free energy. Since $\xi=-\frac{1}{2}\ln(-z)$, the roots in $\xi$-plane have periodically repeating imaginary parts of period $\pi$, and therefore, we choose $-\pi/2\leq\Im\{\xi\}\leq\pi/2$. We fix the real part based on finite-size roots of the AFM mean-field polynomial for a small size. For every $\xi$ in the grid, we set a region in the $m$-plane within which we look for solutions of the saddle point equation using the SHGO algorithm. As explained below Eq.~(\ref{PFSaddles}), only saddle points within the branch cuts of $\Tilde{f}_{\text{AFM}}(m)$ should be considered. The function $\Tilde{f}_{\text{AFM}}(m)$ has two branch cuts at $m=\frac{1}{2k}(\pm \pi + \arg{z})$. Therefore, we set the range of the imaginary part of $m$ to be $\frac{1}{2k}(- \pi + \arg{z})< \Im\{m\} < \frac{1}{2k}(\pi + \arg{z})$. For the real part, by inspection, we found that the saddle points lie in two regions that are $ -2< \Re\{m\} \leq 0$  and $\frac{-\Re\{\xi\}}{2k}-2< \Re\{m\} \leq \frac{-\Re\{\xi\}}{2k}+2$ for the $k$'s we tested.  

\begin{figure}[t!]
    \centering
    \subfigure[$~k=0.5$]{
    \includegraphics[width=0.23\textwidth]{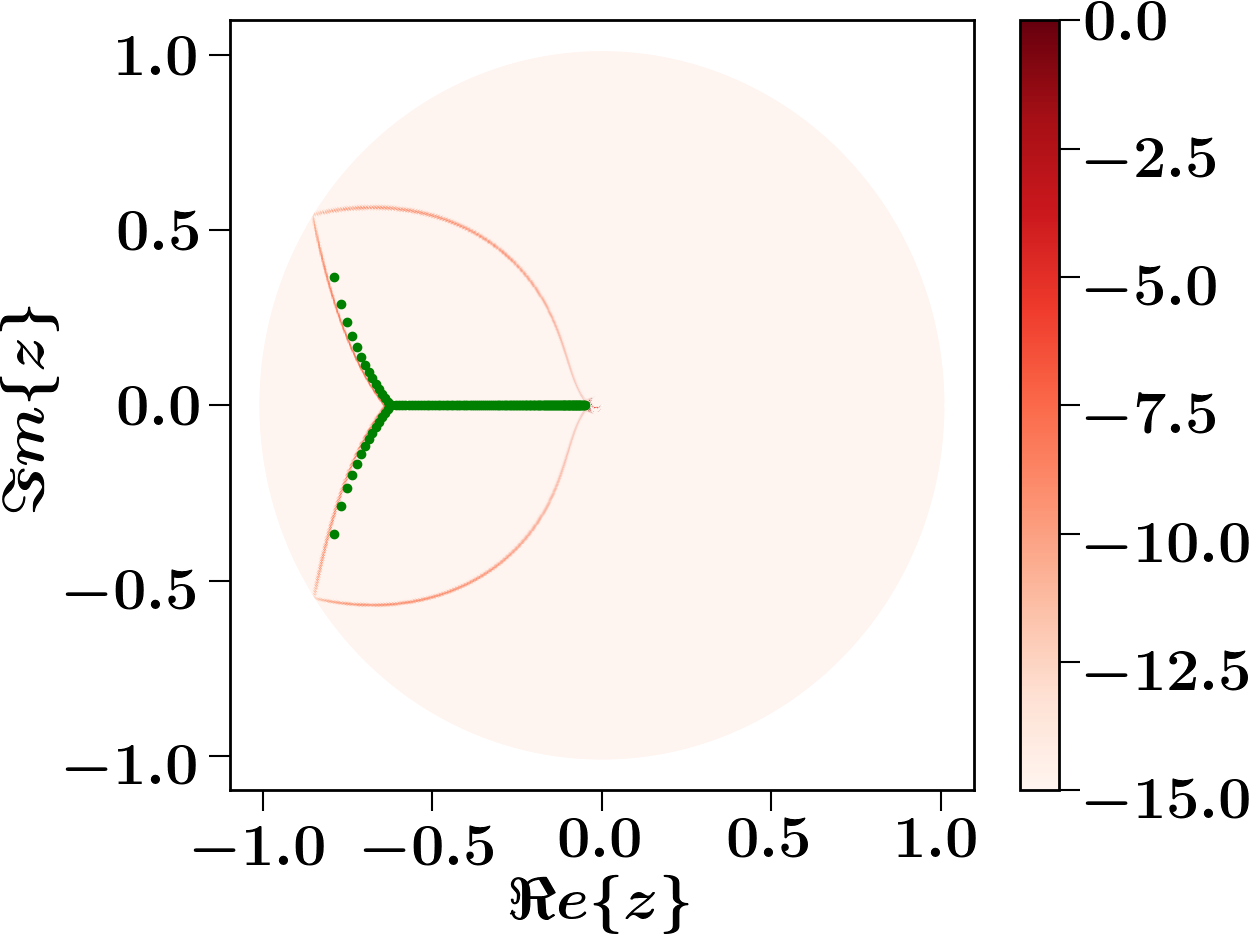}}
    \subfigure[$~k=1$]{
    \includegraphics[width=0.23\textwidth]{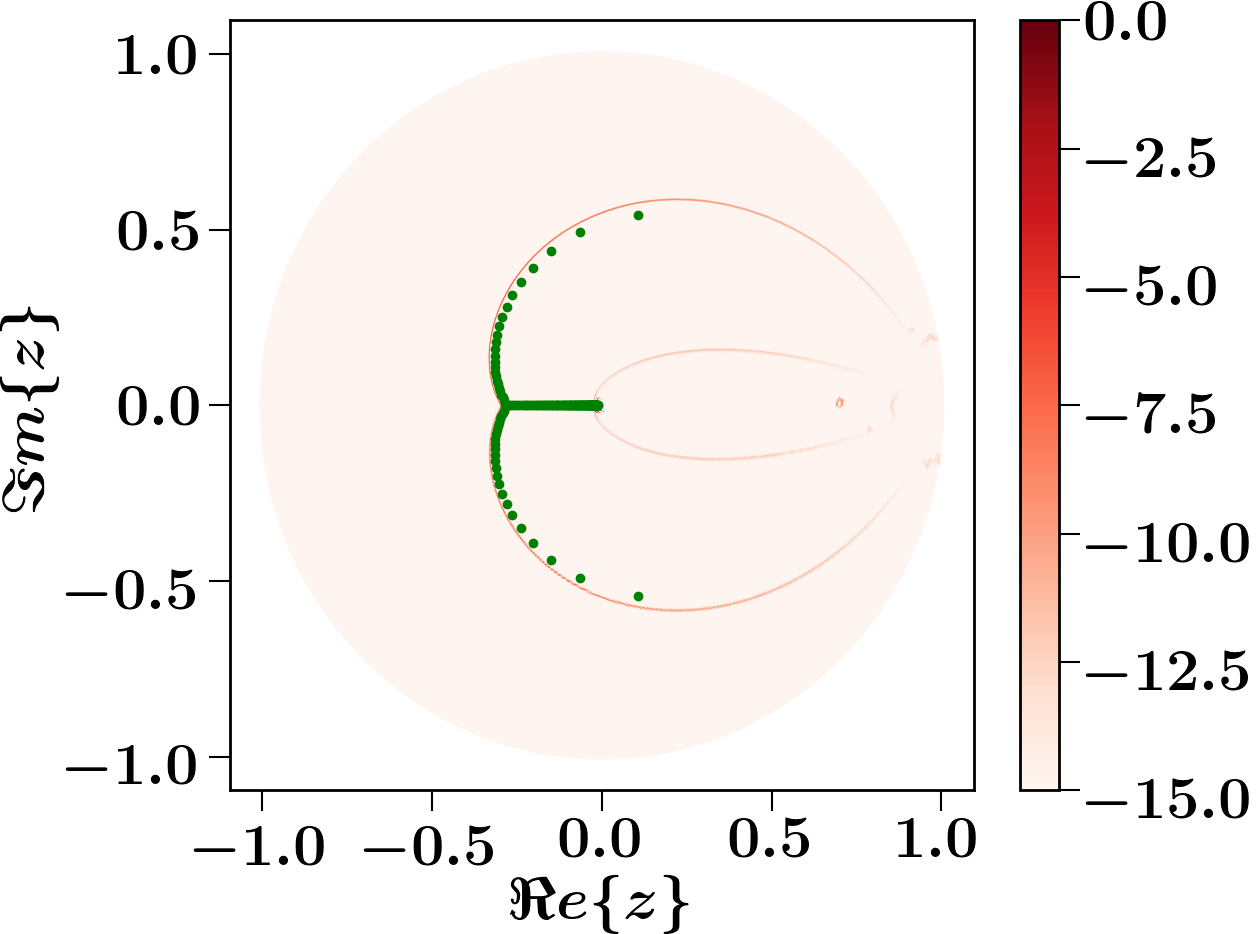}}
    \subfigure[$~k=1.5$]{
    \includegraphics[width=0.23\textwidth]{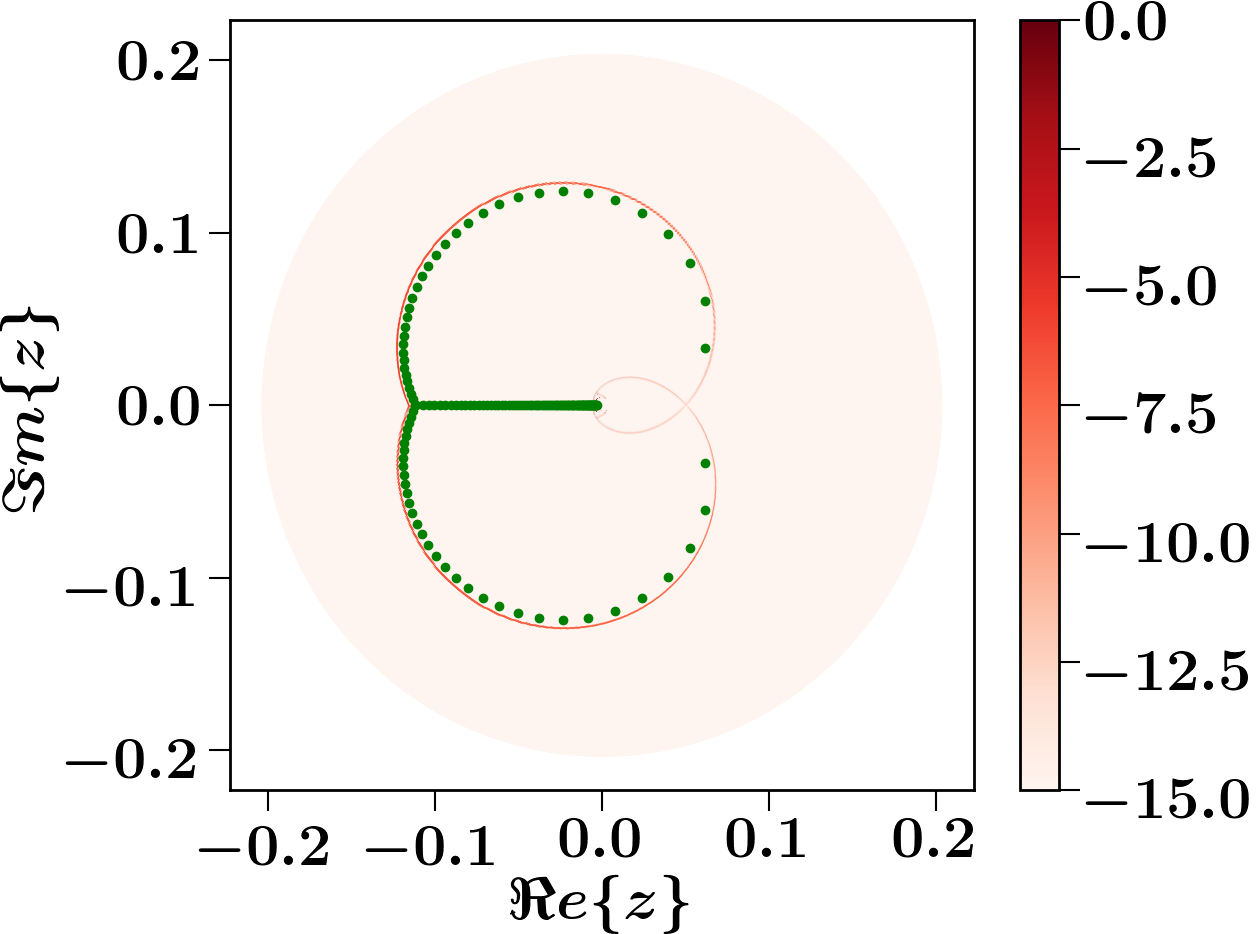}}
    \subfigure[$~k=5$]{
    \includegraphics[width=0.23\textwidth]{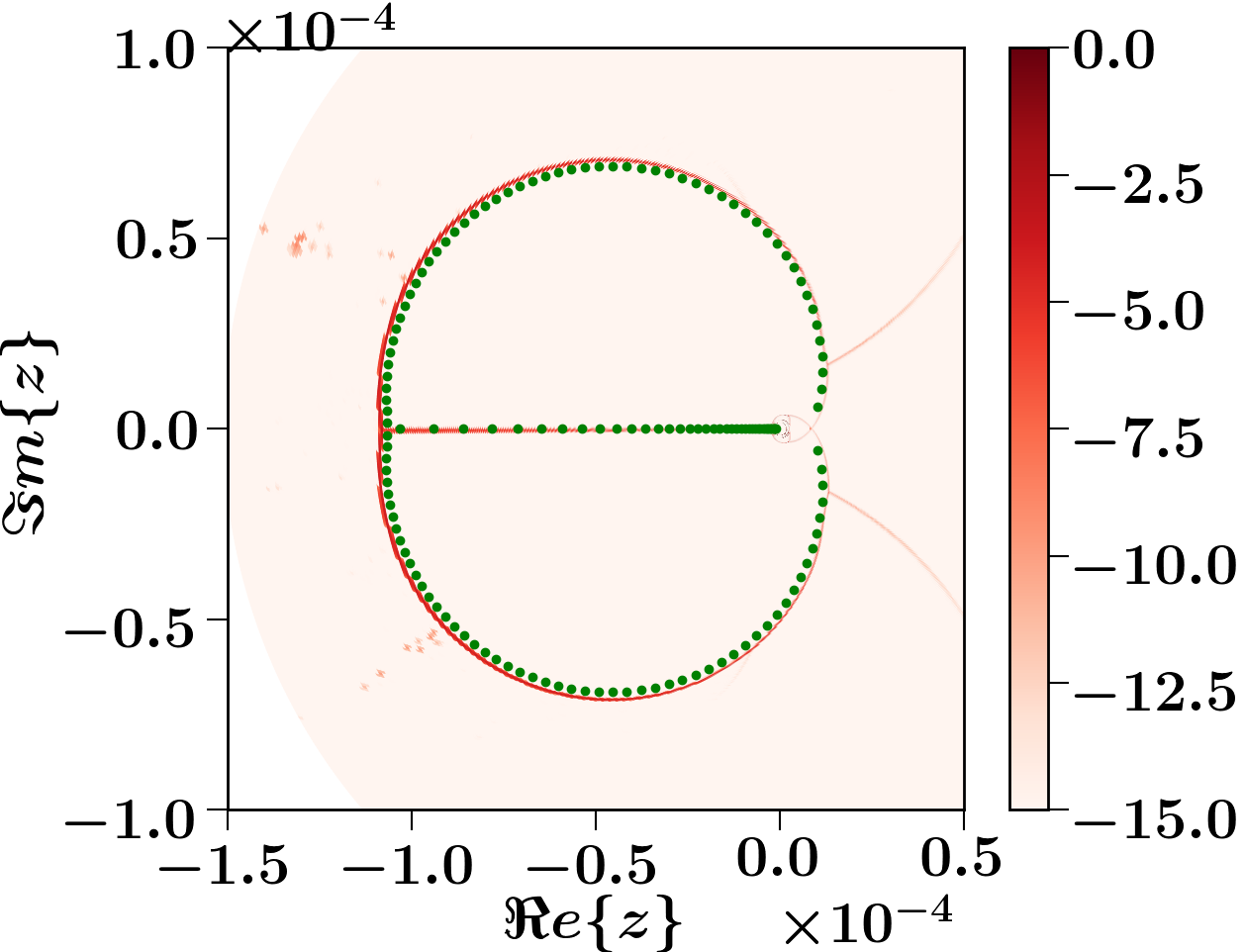}}
    \caption{Contour plot of $\ln{\left|\frac{\rho(z)}{\max\{\rho(z)\}}\right|}$ in $z$-plane at different values of $k$. The exact roots for $N_s=300$ are shown in green. At temperatures greater than the critical temperature (a), the curves are far from the positive real axis. The curve closes onto the positive real axis at $k=k_c$ (b), which marks the phase transition point as the pinching point corresponds to a real value of the magnetic field, namely $z=1\implies h=0$. For lower temperatures $k>k_c$ (c) and (d), the pinching point shifts along the positive real axis with an increasing $|\Re\{h\}|=|\ln{|z|}|$. The critical temperature is $k_c=1$.}
    \label{fig:zroots}
\end{figure}

 Fig.~\ref{fig:xiroots} shows the density of roots obtained in the $\xi$-plane for different values of $k$. These plots indicate that in the thermodynamic limit, the roots form continuous curves as the density is zero everywhere except for the thin 'reddish' curves. The green dots on these figures represent the roots of the finite size mean-field polynomial given by Eq.~(\ref{AFMMeanPoly}) with $N_s=300$. We see that these roots lie exactly on a part of the root curves indicated by these plots. As expected, at temperatures greater than the critical temperature~(see Fig.~\ref{fig:xiroots}a), the curves in the $\xi$-plane do not touch the line $\Im\{\xi\}=\pi/2$ and therefore all the roots are complex. The roots start pinching the line $\Im\{\xi\}=\pi/2$ at $k=k_c=1$. This marks the phase transition point as the pinching point corresponds to a real value of magnetic field $h$, where the partition function vanishes. On lowering the temperature below, $k>k_c$, the pinching point  shifts along the line $\Im\{\xi\}=\pi/2$ with an increasing $\Re\{\xi\}$. This means that the phase transition point at lower temperatures appears at higher values of $|h|$. Similar features are seen when the roots are plotted in $z$-plane and the results are shown in Fig.~\ref{fig:zroots}.

A part of the root density curves in Fig.~\ref{fig:xiroots} does not have corresponding $N_s=300$ roots (green dots). A natural question is
to ask whether these are indeed root curves of the partition function or not. To answer this question, consider two saddle points, $m_1$ and $m_2$, with the lowest real part of free energy for a given $\xi$ and $k$. After saddle point approximation, the partition function is 
\begin{equation}
    \mathcal{Z}_{\text{AFM}}\sim e^{-N_s\beta f(m)}\left(1+e^{-N_s \beta \left(f(m')-f(m)\right)}\right),
    \label{ZAFMfinal}
\end{equation}
 where $f(m)=\tilde{f}_{\text{AFM}}(m,\tanh(-km+\beta h))$, and $m$($m'$) is such that $\Re\{f(m)\}$($\Re\{f(m')\}$) is the minimum(maximum) between $\Re\{f(m_1)\}$ and $\Re\{f(m_2)\}$. The partition function then vanishes under the condition $\beta \left(f(m')-f(m)\right)= \pi i (2n+1)/N_s$. This implies that the roots happen when $\Re\{f(m')\}=\Re\{f(m)\}$, and $\Im\{f(m')\}-\Im\{f(m)\}=\pi (2n+1)/N_s$ for an integer $0\leq n < N_s$, given finite $N_s$. As $N_s\rightarrow \infty$, the condition on the imaginary parts is relaxed because all integers $n$ are allowed. Hence, any different imaginary parts may result in a root. This is equivalent to the criteria developed by Ohminami et. al. in~\cite{ohminami1972distribution}.  

\begin{figure}[t!]
    \centering
    \includegraphics[width=0.5\textwidth]{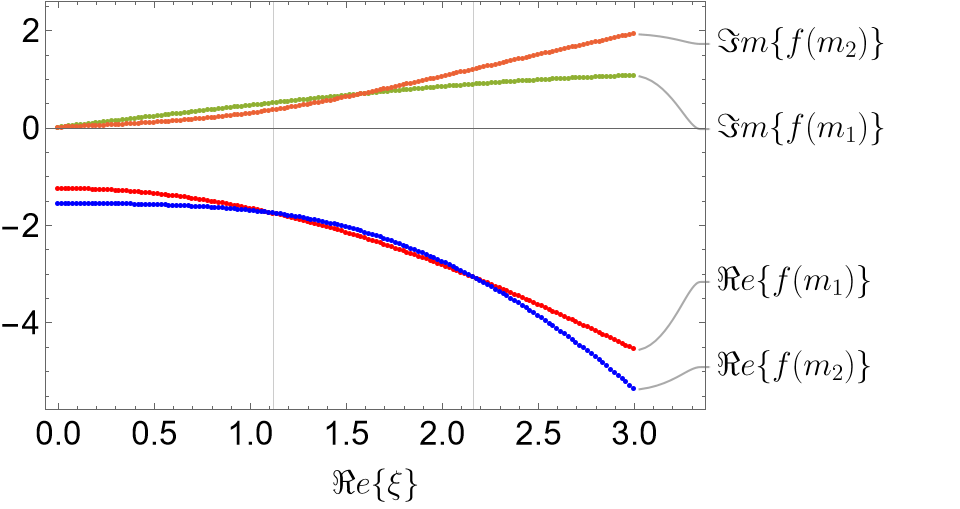}
    \caption{The variation of the real and imaginary parts of the free energies $f_{\text{AFM}}(m_1)$ and $f_{\text{AFM}}(m_2)$ with the $\Re\{\xi\}$ at a fixed $\Im\{\xi\}=1$ and at $k=1.5$. $m_1$ and $m_2$ are the two saddle points with the lowest real parts of free energy. The two vertical grid lines are the values of $\Re\{\xi\}$ at which the roots occur. At those lines, the free energies at the two saddle points have the same real part but different imaginary parts resulting in a vanishing expression of the partition function.}
    \label{fig:SameReFe}
\end{figure}

To test the above condition, we study the free energy as a function of $\Re\{\xi\}$, with $\Im\{\xi\}=1$ at $k=1.5$. The free energy of the two saddle points $m_1$ and $m_2$ is plotted in Fig.~\ref{fig:SameReFe}.
The two vertical grid-lines are at the values of $\Re\{\xi\}$ at which a line $\Im \{\xi\}=1$ cuts the root curves in Fig.~\ref{fig:xiroots}c. In between the vertical grid-lines, which corresponds to the region inside the root curves in Fig.~\ref{fig:xiroots}c, $\Re\{f(m_1)\}$ is smaller than $\Re\{f(m_2)\}$ whereas outside the grid-lines, $\Re\{f(m_2)\}$ is smaller. Therefore, in the thermodynamic limit, $f(m_1)$ and $f(m_2)$ approximate the free energy inside and outside of the vertical grid lines, respectively. At the vertical grid lines, $f(m_1)$ and $f(m_2)$ have the same real parts but different imaginary parts. This implies that both curves are indeed root curves in the thermodynamic limit. 

\begin{figure}[t!]
    \centering
    \subfigure[$~k=0.5$]{
    \includegraphics[width=0.22\textwidth]{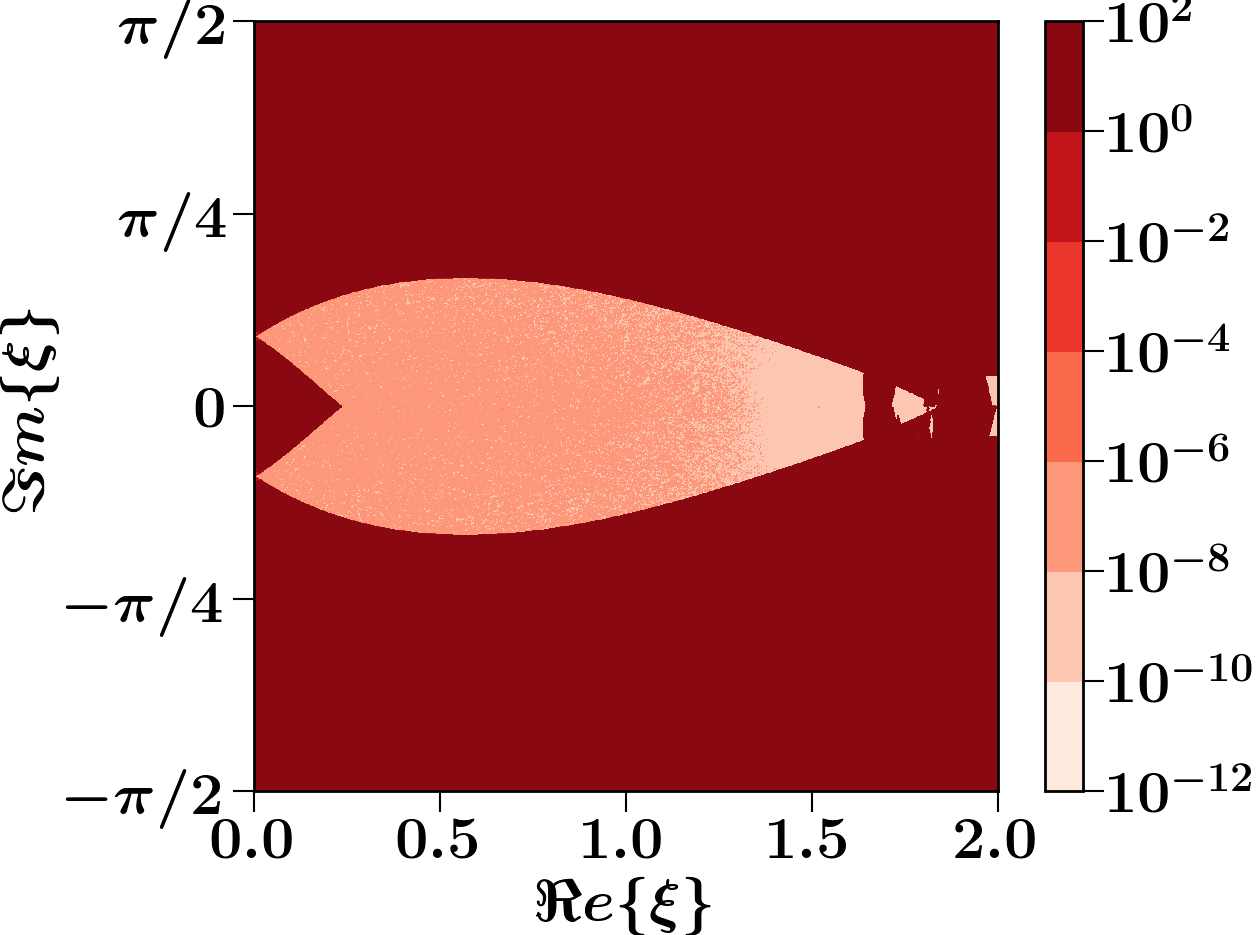}}
    \subfigure[$~k=1$]{
    \includegraphics[width=0.22\textwidth]{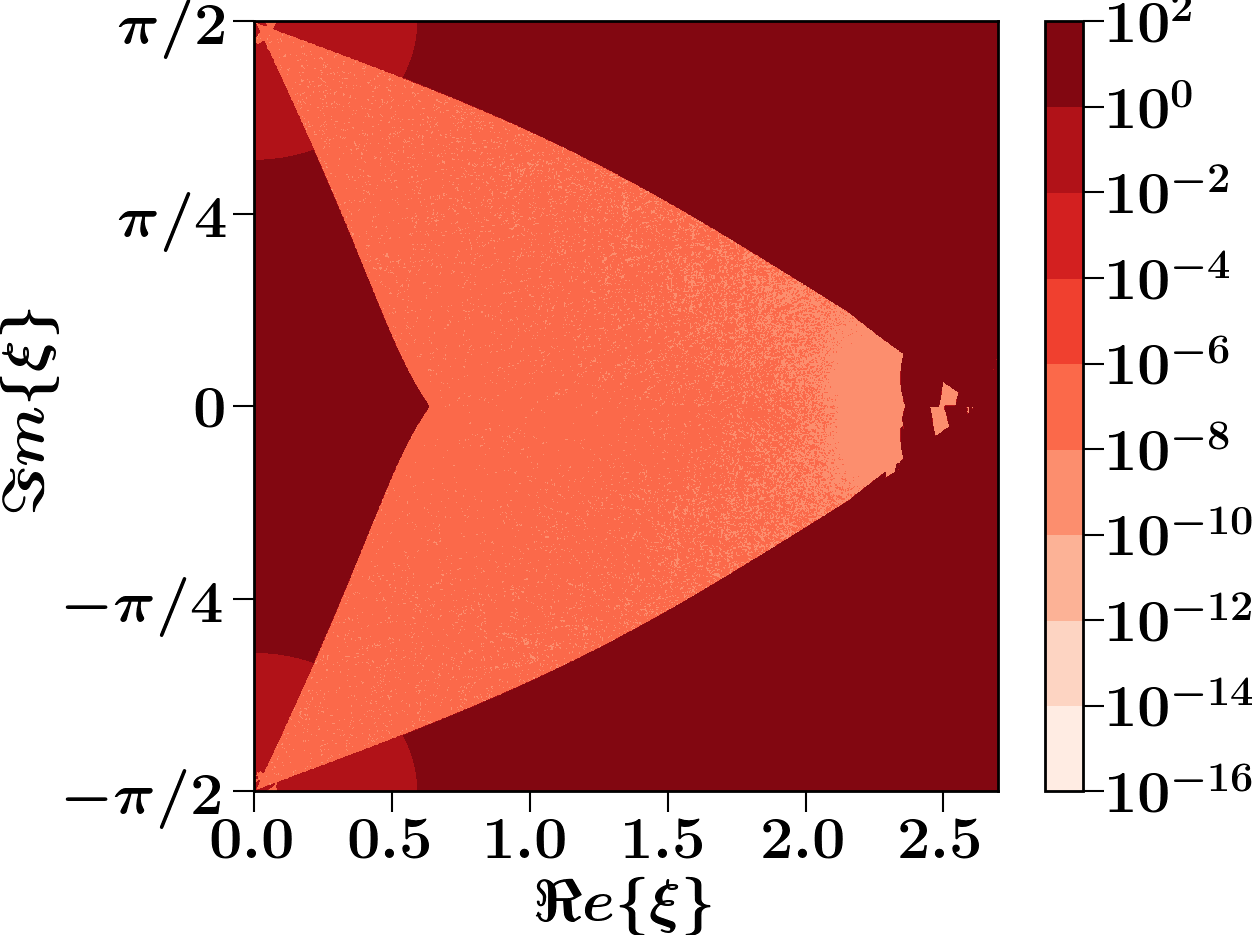}}
    \subfigure[$~k=1.5$]{
    \includegraphics[width=0.22\textwidth]{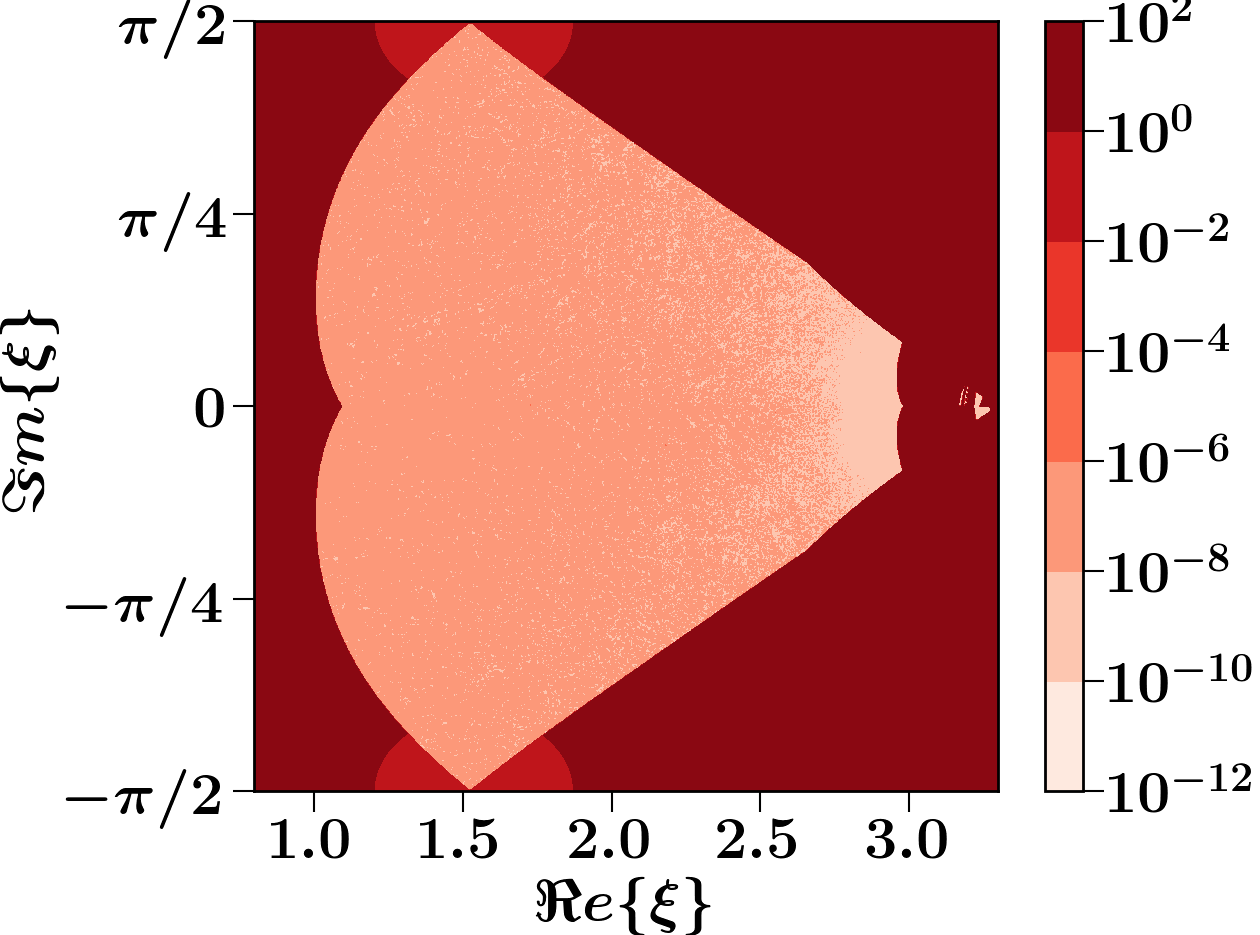}}
    \subfigure[$~k=5$]{
    \includegraphics[width=0.22\textwidth]{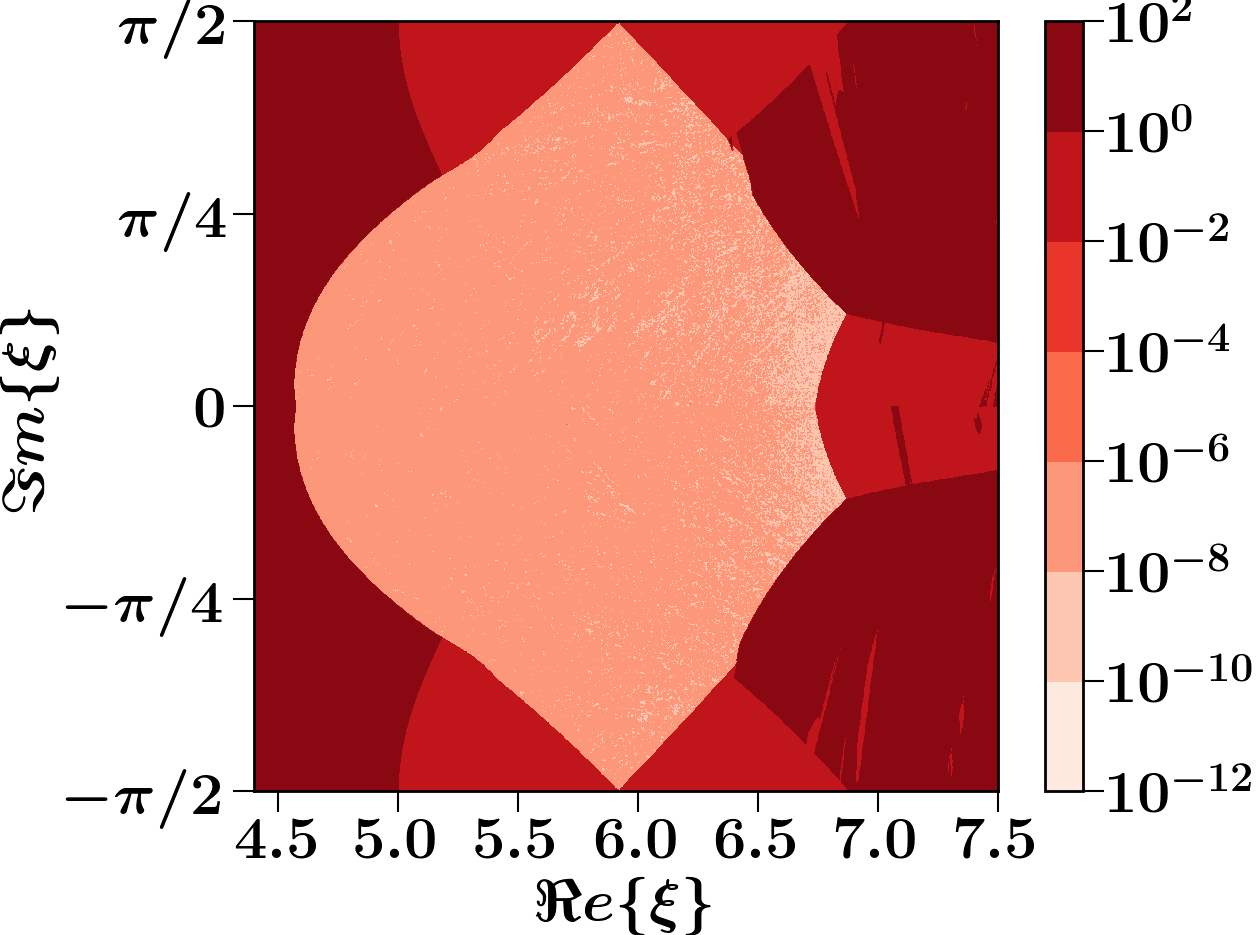}}
    \caption{Contour plot of the magnitude of the staggered magnetization $m_s=\left|m_a^*-m_b^*\right|$ in $\xi$-plane at different values of $k$, where the critical temperature at $k_c=1$ and the legend bar is given in $\log$ scale. For all temperatures, the staggered magnetization vanishes in the region bounded by the root curves, while it is finite in the outside region showing the existence of complex paramagnetic and antiferromagnetic phases.}
    \label{fig:exOrderContour}
\end{figure}
\begin{figure}[t!]
    \centering
    \subfigure[$~k=0.5$]{
    \includegraphics[width=0.22\textwidth]{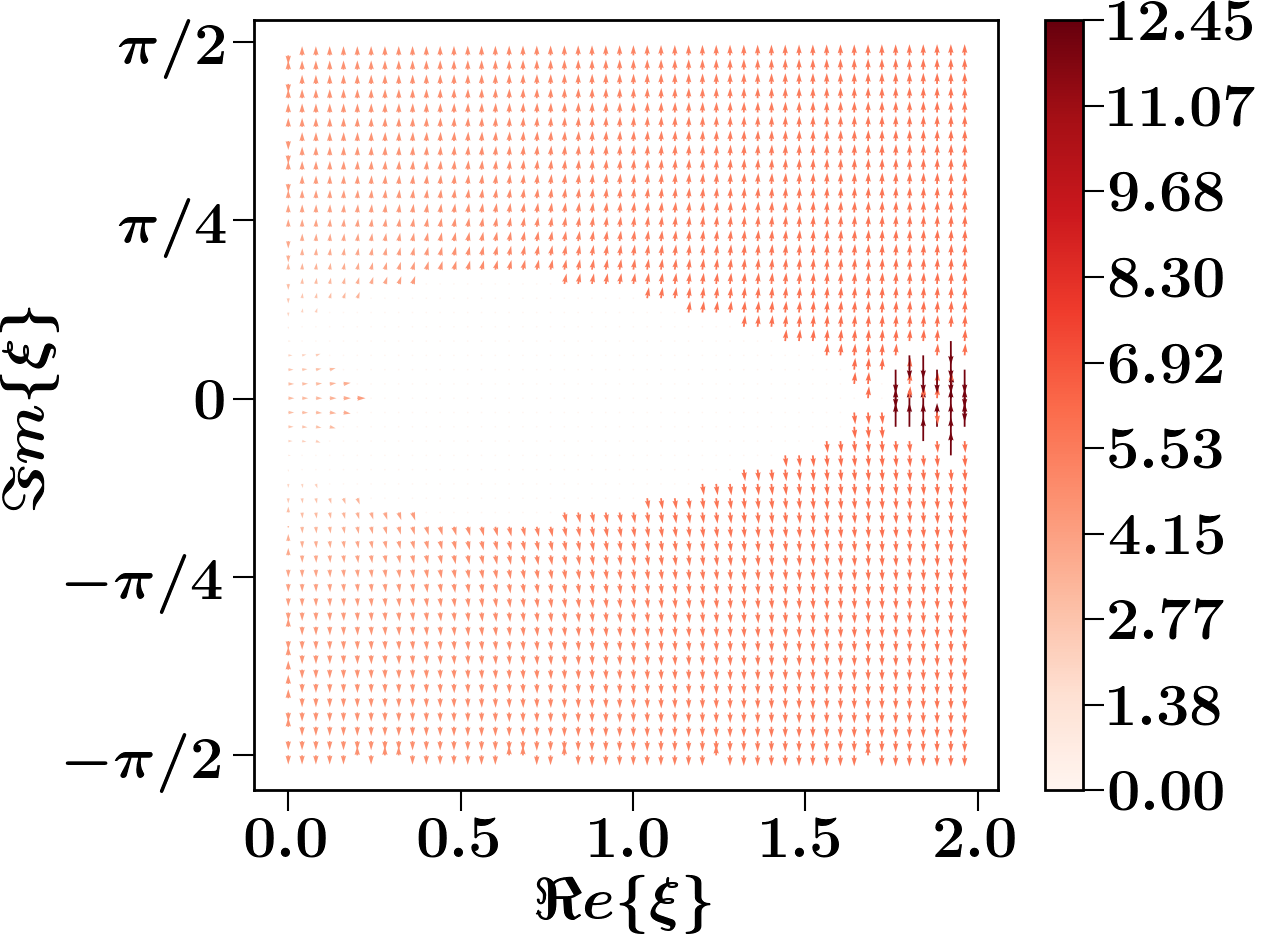}}
    \subfigure[$~k=1$]{
    \includegraphics[width=0.22\textwidth]{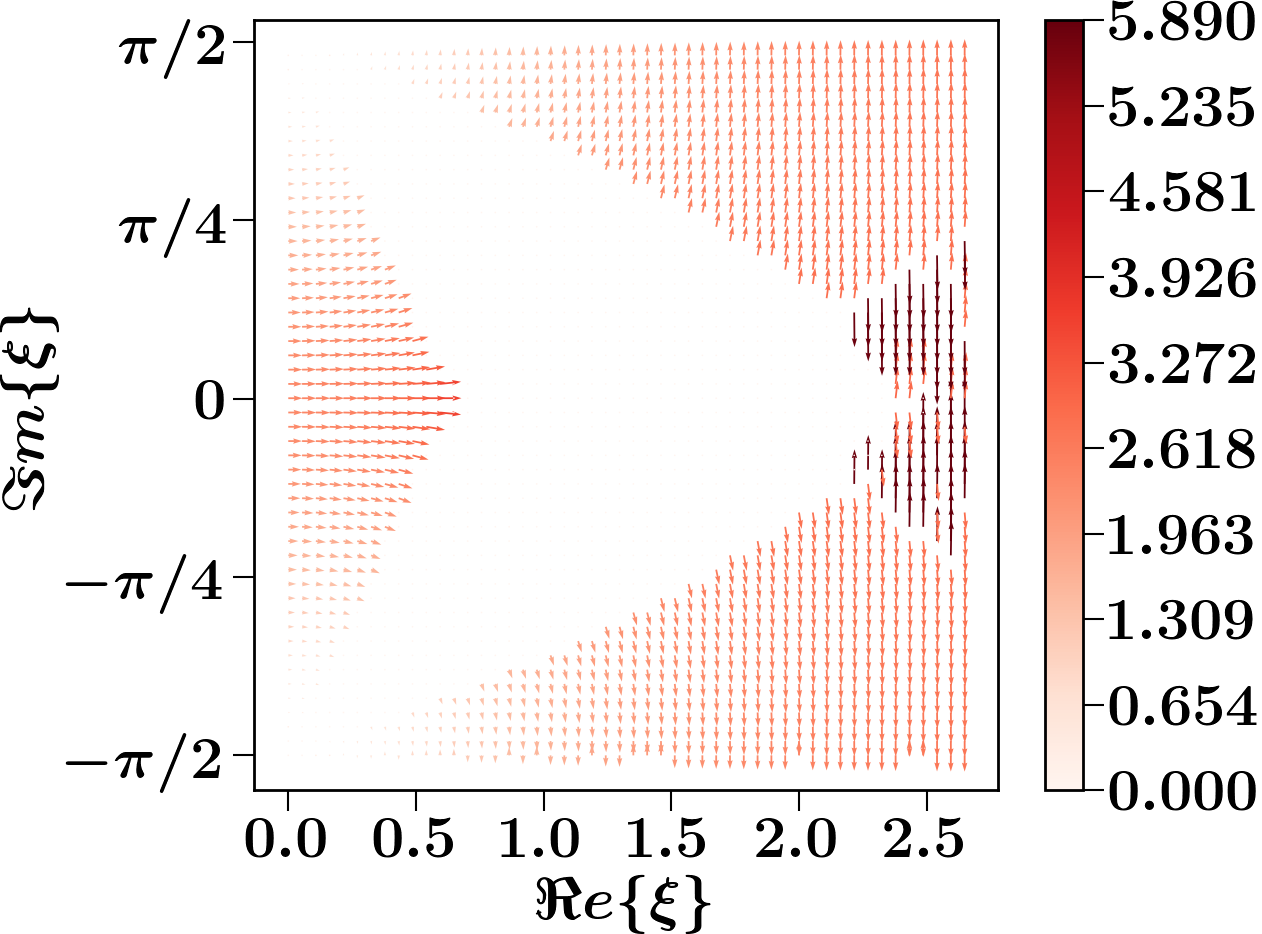}}
    \subfigure[$~k=1.5$]{
    \includegraphics[width=0.22\textwidth]{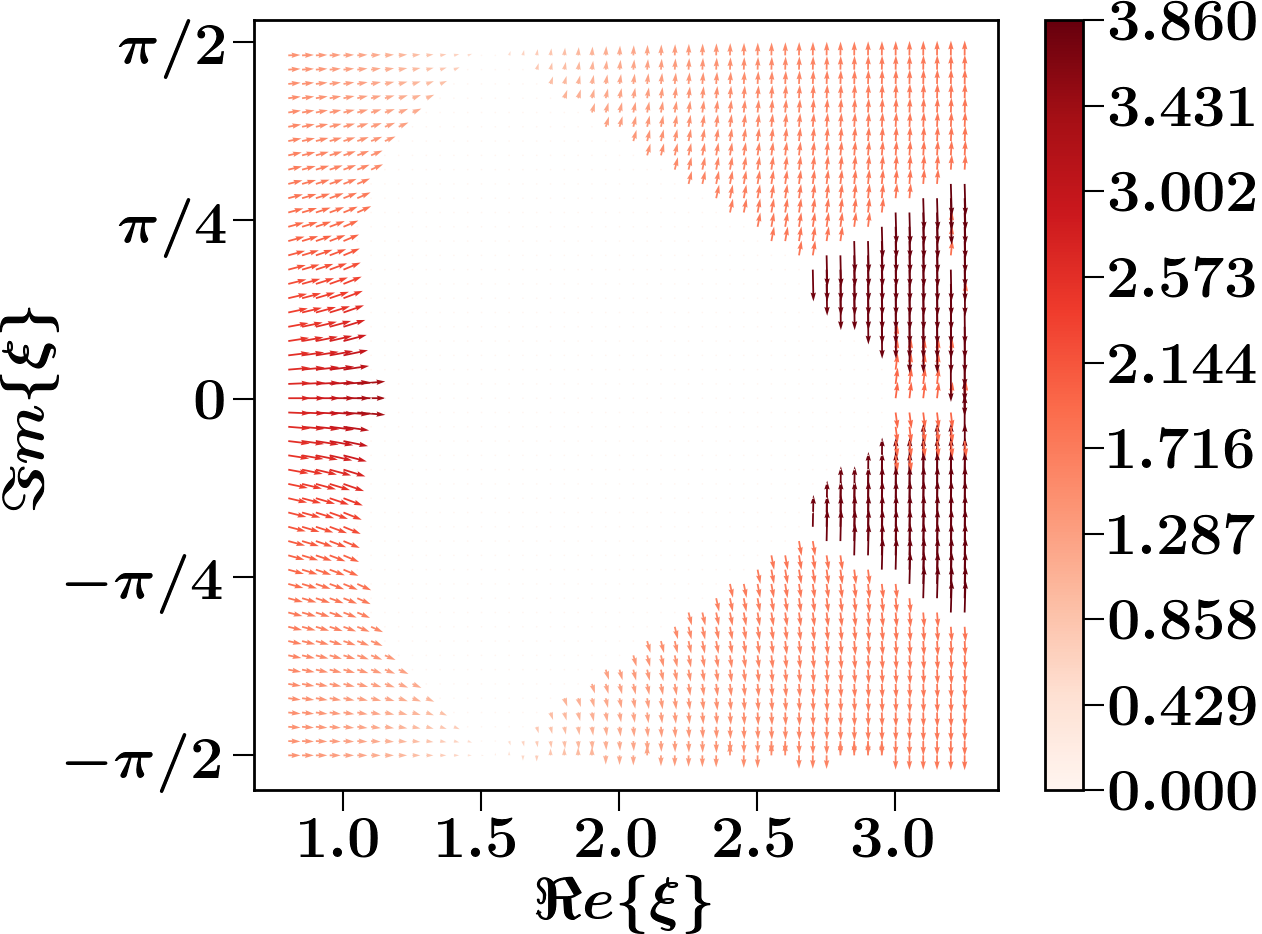}}
    \subfigure[$~k=5$]{
    \includegraphics[width=0.22\textwidth]{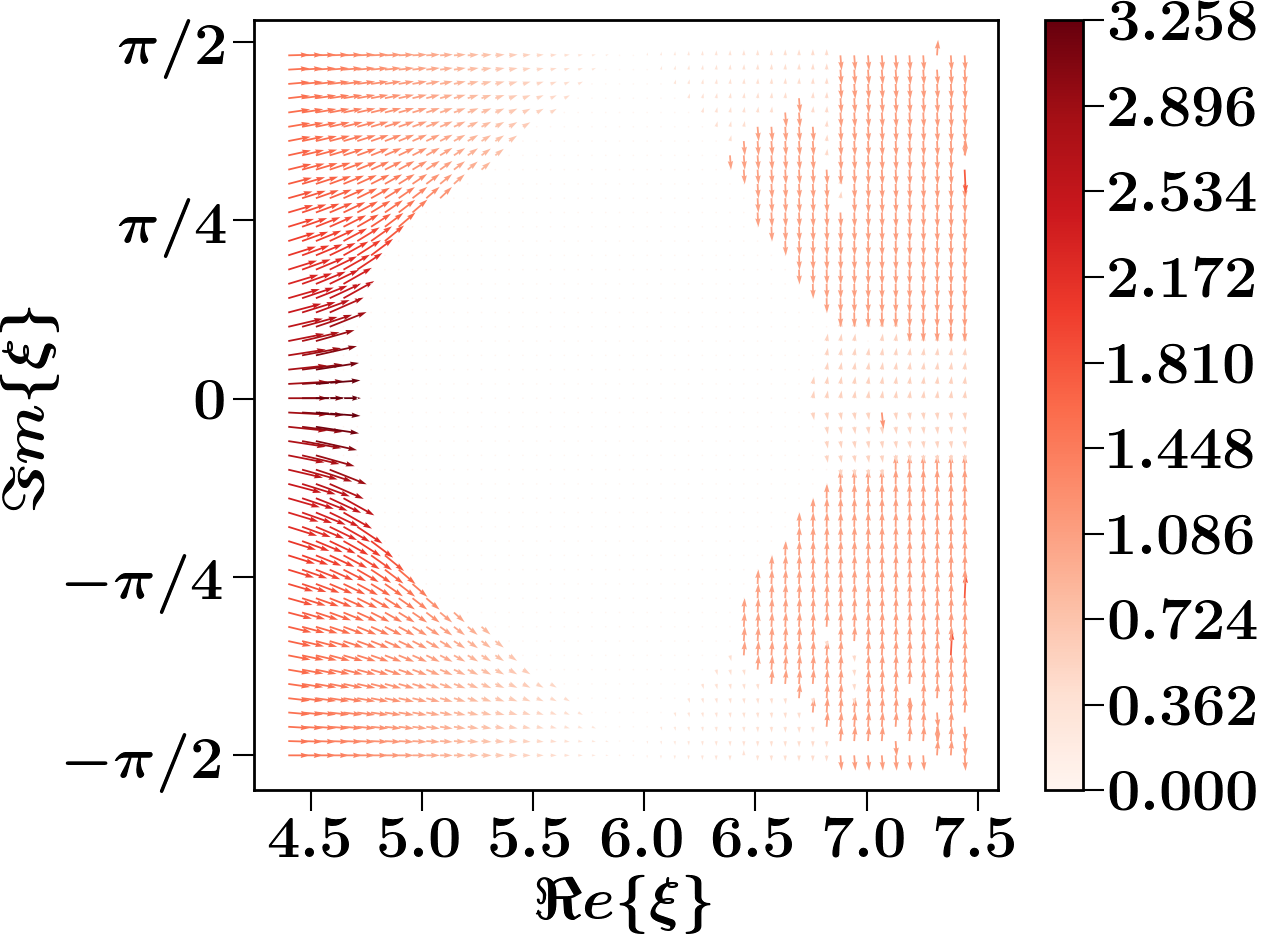}}
    \caption{Vector plot of the staggered magnetization, where at each point in $\xi$-plane, the vector plotted is $\Re\{m_a^*-m_b^*\} \hat{x}+\Im\{m_a^*-m_b^*\} \hat{y}$ at different values of $k$, where the critical temperature at $k_c=1$ and the legend bar indicates the magnitude of the vector. For all temperatures, the vectors have zero magnitude inside the region bounded by the root curves (complex paramagnetic phase) and finite magnitude outside (complex antiferromagnetic phase).}
    \label{fig:exOrderVec}
\end{figure}

It is also interesting to look at the nature of the solution, $m_a^*$  and $m_b^*$, of the saddle point equations Eqs.~(\ref{AFMsaddleEqs1},\ref{AFMsaddleEqs2}), which minimize the free energy. In Fig.~\ref{fig:exOrderContour}, we plot the absolute value of the staggered magnetization, $|m_a^*-m_b^*|$. We observe that the regions bounded by the root curves have vanishing staggered magnetization, while in the outside region, the staggered magnetization is finite. Therefore, the root curves demarcate the $\xi$-plane into regions of zero and non-zero staggered magnetization. This shows two different phases that exist in the complex $\xi$-plane: a complex paramagnetic phase inside the root curves and a complex anti-ferromagnetic phase outside of the root curves. We also see this in the plot of the staggered magnetization as a vector field in $\xi$-plane by plotting the vector $\Re\{m_a^*-m_b^*\} \hat{x}+\Im\{m_a^*-m_b^*\} \hat{y}$ at every point. Fig.~\ref{fig:exOrderVec} shows this vector plot of the staggered magnetization for different values of $k$. Note that the $\Re\{m_a^*-m_b^*\}$ vanishes for $\Re\{\xi\}$ greater than the pinching point (critical magnetic field) as expected from the mean-field AFM phase diagram. We see that the vectors have zero magnitude inside the region bounded by the root curves and finite magnitude outside confirming the views of two different phases in the complex $\xi$-plane.

\section{Conclusion}
In summary, we have shown analytically that the logarithm of Yang-Lee zeros of the nearest neighbor Ising model scale as $\sqrt{k}$ at high temperatures for an arbitrary regular lattice in any dimension. Assuming these logarithms have a power series at high temperatures around $k=0$, we find novel constraints (i.e. sum rules) for the coefficients of the power series. In general, these sum rules depend on the number of sites, bonds, and the boundary conditions of the lattice. However, we show here that a linear combination of these sum rules is independent of the boundary conditions. We verified the sum rules for two cases (i) the 1-d nearest neighbor Ising model with periodic boundary conditions, and (ii) the  2-d square lattice with open boundary conditions. For the latter, the verification was done by computing the exact partition function up to size $16\times16$, which seems to be the largest size studied for the purpose of computing the partition function zeros. We also examined the behavior of Yang-Lee zeros at different values of the temperature for this lattice size. 

To understand the behavior of the Yang-Lee zeros in the thermodynamic limit, we studied the mean-field model with infinite-ranged FM and AFM coupling. We showed that the logarithm of Yang-Lee zeros of this model also scale as $\sqrt{k}$ at high temperatures using two different methods. The first method followed the same power series expansion as for the Ising model. The second method was by showing that the AFM mean-field partition function at high temperatures is a linear combination of Hermite polynomials, which leads to the roots having the same $k$ dependence at high temperatures as the Ising model. We then studied the roots of the FM and AFM partition functions of the mean-field model in the thermodynamic limit using a simple but powerful approach involving the mean-field free energy. Using this approach, and useful newly developed iterative schemes,  we numerically determined the root curves for the FM and AFM cases. For the AFM case, our results show new root curves that, to our knowledge, were not reported in earlier literature. 
For the largest size we computed, none of the roots were in the vicinity of the new root curve. Therefore, the roots at finite sizes, may not reflect the entire locus of zeros in the thermodynamic limit. Summarizing, we find that the new root curves demarcate the boundary between distinct "thermodynamic phases", characterized by different (complex) values of the staggered magnetization. This provides a helpful physical basis for understanding the root-curves.

\section{Acknowledgments}
The work at UCSC was supported by the US Department of Energy (DOE), Office of Science, Basic
Energy Sciences (BES), under Award No. DE-FG02-06ER46319.

\clearpage
 \bibliography{biblio}
 \bibliographystyle{ieeetr}

\appendix
\section{Exact Algorithm}
\label{Appx:Bhanot}
It is computationally difficult to compute the exact partition function by directly enumerating all the configurations of the lattice spins due to memory limitations. Binder~\cite{binder1972statistical} introduced a memory-efficient algorithm to compute the zero-field partition function of the nearest-neighbor Ising Hamiltonian that enumerates all configurations of the spins on a $d$-dimensional lattice by iteratively building the lattice from the partition function of the $(d-1)$-dimensional lattice. The algorithm stores the coefficients of a polynomial in the variable $u$, hence, it still has a memory limitation that scales with the degree of the polynomial~\cite{bhanot1990numerical}. Bhanot proposed a modification of Binder's algorithm~\cite{bhanot1990numerical} to overcome this memory limitation. Bhanot's trick is to assume that $u^{m_u}=c_u$ for an integer $m_u<N_b$ and a real number $c_u$. This reduces the degree of the polynomial to $m_u-1$, consequently, reducing the memory needed to store the coefficients. A process that Bhanot called “folding” the polynomial, where the coefficients of the original polynomial could be retrieved from the coefficient of the folded polynomial by knowing the values $m_u$ and $c_u$.\\

In the presence of a magnetic field, the partition function is a polynomial of two variables $u$ and $z$ as shown in Eq.~(\ref{Z(omega)}). To adapt Bhanot's technique for a polynomial of two variables, a similar trick is used for the new variable $z^{m_z}=c_z$ for an integer $m_z<N_s$ and a real number $c_z$. The partition function becomes
\begin{align}
\mathcal{Z}&=\sum_{n_b=0}^{N_b}\sum_{n_s=0}^{N_s}\Omega(n_b,n_s)u^{n_b}z^{n_s}\notag\\
&=\sum_{n_b=0}^{m_u-1}\sum_{n_s=0}^{m_z-1}\Tilde{\Omega}_{m_u,m_z}(n_b,n_s)u^{n_b}z^{n_s}.
    \label{Zfolded}
\end{align} The relation between the coefficients of the original and folded polynomial is given by 
\begin{align}
    \Tilde{\Omega}_{m_u,m_z}&(n_b,n_s)\notag\\
    &=\sum_{\mu=0}^{\lfloor{\frac{N_b-n_b}{m_u}}\rfloor}\sum_{\nu=0}^{\lfloor{\frac{N_s-n_s}{m_z}}\rfloor}\Omega(n_b+\mu m_u,n_s+\nu m_z)c_u^\mu c_z^\nu.
    \label{OmegaRetriveGeneral}
\end{align}
For every choice of $m_u$ and $m_z$, we get a different set of $m_u\times m_z$ independent equations that are linear in $\Omega(n_b,n_s)$ from Eq.~(\ref{OmegaRetriveGeneral}).
In principle, $(N_b+1)(N_s+1)$ independent equations are needed to determine all original coefficients $\Omega(n_b,n_s)$. Running Bhanot's algorithm for different values of $c_u$ and $c_z$ in order to generate the required number of independent equations, then solving the linear system completely determines the coefficients of the partition function in Eq.~(\ref{Z(omega)}). 

For the Ising model on an $L\times L$ lattice, it is enough to choose $m_z=\lfloor{\frac{N_s}{2}}\rfloor+1$ and make use of the symmetry in the coefficients of $z$ due to the symmetry $\sigma_i \rightarrow -\sigma_i$. With this choice, it is enough to set $c_z=0$. We then set $m_u=L$ and vary $c_u$ for different integer values until $(N_b+1)$ equations are generated. The number of  different values of $c_u$ is $\lfloor{\frac{N_b+1}{m_u}}\rfloor+1$. \\

\section{AFM MF Polynomial as Hermite Polynomials}
\label{Appx:HermiteDerivation}

To write the polynomial in an integral form, we start by using the identity 
\begin{align}
    &e^{-\frac{1}{2}N_skm_am_b}=e^{\frac{1}{2}N_sk(m_s^2-m_t^2)}\notag\\
    &\quad \quad=\frac{N_s}{2\pi k}\int_{-\infty}^{\infty}\int_{-\infty}^{\infty}\mathrm{d}x\mathrm{d}ye^{-\frac{N_s}{2k}(x^2+y^2)}e^{N_s(xm_s+iym_t)},
    \label{GaussInteg}
\end{align}
where $m_s=m_a-m_b$ is the staggered magnetization, and $m_t=m_a+m_b$ is the total magnetization. Now, expressing $m_s$ and $m_t$ in terms of $M_a$ and $M_b$ will allow us to carry out the summations in the polynomial. We note that 
\begin{equation}
    N_s(xm_s+iym_t)=2M_a(x+iy)+2M_b(-x+iy)-iyN_s.
    \label{msmt}
\end{equation}
Plugging this back into Eq.~(\ref{AFMMeanPoly}) and carrying out the two sums using binomial expansion, we reach the polynomial in the integral form  
\begin{align}
    \mathcal{Z}_{\text{AFM}}=\frac{N_s}{2\pi k}&\int_{-\infty}^{\infty}\int_{-\infty}^{\infty}\mathrm{d}x\mathrm{d}ye^{-\frac{N_s}{2k}(x^2+y^2)-iN_sy}\notag\\
    &\times\left(1+ze^{2(x+iy)}\right)^\frac{N_s}{2}\left(1+ze^{2(-x+iy)}\right)^\frac{N_s}{2}.
    \label{AFMIntegral}
\end{align}
We now use $z=-e^{-2\xi}$ to simplify the parentheses in terms of hyperbolic functions as 
\begin{align}
    \mathcal{Z}_{\text{AFM}}=\frac{2^{N_s}N_s}{2\pi k}&e^{-N_s\xi}\int_{-\infty}^{\infty}\int_{-\infty}^{\infty}\mathrm{d}x\mathrm{d}ye^{-\frac{N_s}{2k}(x^2+y^2)}\notag\\
    &\times\bigg(\sinh{\left(x+iy-\xi\right)}\sinh{\left(-x+iy-\xi\right)}\bigg)^\frac{N_s}{2}.
    \label{AFMIntegral2}
\end{align}
The high temperature limit is $k\rightarrow 0$, we therefore make the scaling transformation $(\Tilde{x},\Tilde{y},\Tilde{\xi})\rightarrow (\alpha x,\alpha y,\alpha \xi)$, where $\alpha=\sqrt{\frac{2k}{N_s}}$. Since the high temperature limit now is $\alpha \rightarrow 0$, we then use the expansion $\sinh({\alpha A})\approx \alpha A$ to get 
\begin{align}
 \mathcal{Z}_{\text{AFM}}\approx\frac{(2\alpha)^{N_2}}{\pi}e^{-\alpha N_s\tilde{\xi}}\int_{-\infty}^{\infty}\int_{-\infty}^{\infty}&\mathrm{d}\tilde{x}\mathrm{d}\tilde{y}e^{-(\tilde{x}^2+\tilde{y}^2)}\notag\\
 &\times\left((i\tilde{y}-\tilde{\xi})^2-\tilde{x}^2\right)^\frac{N_s}{2}.
    \label{AFMIntegApprox}
\end{align}
The integration over $\tilde{x}$ can be evaluated after expanding the parentheses using binomial expansion  
\begin{align}
    \sum_{p=0}^{N_s/2}\binom{\frac{N_s}{2}}{p}(-1)^{p}&\int_{-\infty}^{\infty}\mathrm{d}\tilde{x}e^{-\tilde{x}^2}\tilde{x}^{2p}\notag\\
    &=\sum_{p=0}^{N_s/2}\binom{\frac{N_s}{2}}{p}(-1)^{p}\sqrt{\pi}\frac{(2p)!}{2^{2p}p!}.
    \label{xintegral}
\end{align}

\begin{figure}[t!]
    \centering
    \subfigure[$~k=10^{-1}k_c$]{
    \includegraphics[width=0.45\textwidth]{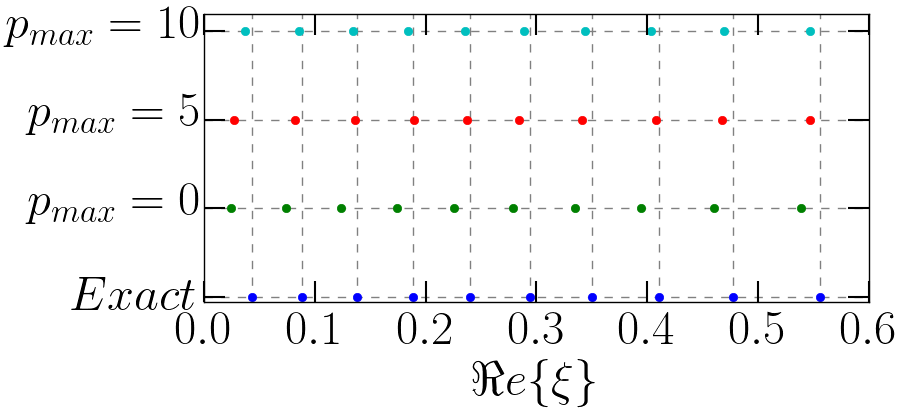}}
    \subfigure[$~k=10^{-3}k_c$]{
    \includegraphics[width=0.45\textwidth]{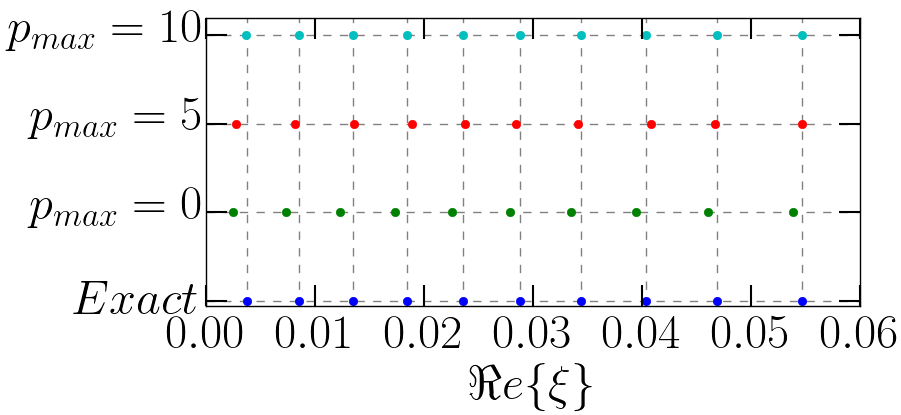}}
    \caption{The roots $\xi_j=\alpha \tilde{\xi}_j$ of the linear combinations of Hermite polynomials in Eq.~(\ref{AFMHermite}) terminated at different values of $p_\text{max}$ for $N_s=20$ and different values of high temperatures. The roots are all real and simple.}
    \label{fig:MFRootsHighTemp}
\end{figure}

Plugging this back into Eq.~(\ref{AFMIntegApprox}), we get 
\begin{align}
    \mathcal{Z}_{\text{AFM}}\approx\alpha^{N_s}e^{-\alpha N_s\tilde{\xi}}&\sum_{p=0}^{N_s/2}\binom{\frac{N_s}{2}}{p}(-1)^{p}\sqrt{\pi}\frac{(2p)!}{p!}\frac{2^{N_s-2p}}{\sqrt{\pi}}\notag\\
    &\times\int_{-\infty}^{\infty}\mathrm{d}\tilde{y}e^{-\tilde{y}^2}(-\tilde{\xi}+i\tilde{y})^{N_s-2p}.
    \label{AFMSingleInt}
\end{align}
Finally, using the definition of Hermite polynomial
\begin{equation}
    H_n(t)=\frac{2^{n}}{\sqrt{\pi}}\int_{-\infty}^{\infty}\mathrm{d}\tilde{y}e^{-\tilde{y}^2}(t+i\tilde{y})^{n},
    \label{HermiteDef}
\end{equation}
the leading behavior of the AFM mean-field polynomial at high temperatures is given by 
\begin{equation}
    \mathcal{Z}_{\text{AFM}}\approx\left(-\alpha e^{-\alpha \tilde{\xi}}\right)^{N_s}\sum_{p=0}^{N_s/2}(-1)^p\binom{\frac{N_s}{2}}{p}\frac{(2p)!}{p!}H_{N_s-2p}(\tilde{\xi}).
    \label{AFMHermiteAppx}
\end{equation}
The function in Eq.~(\ref{AFMHermiteAppx}) is a linear combination of Hermite polynomials in the variable $\Tilde{\xi}$ modulated with the factor $e^{-\alpha N_s\tilde{\xi}}$. Numerical calculations show that the roots of this linear combination of Hermite polynomials are real for different values of $N_s$.
A comparison between the exact roots of $\mathcal{Z}_{\text{AFM}}$ and the roots of this linear combination of Hermite polynomials terminated at some given $p_\text{max}$ for $N_s=20$ and at two temperatures $k=10^{-1}k_c$ and $10^{-3}k_c$ are shown in Fig.~\ref{fig:MFRootsHighTemp}. The figure shows the exact roots on the lowest horizontal level, and the roots for $p_{\text{max}}=0,5,10$ at the higher levels. $p_{\text{max}}=10$, which corresponds to $N_s/2$, gives the best approximation for the exact roots and approximations become better at higher temperatures.

\end{document}